\documentclass[a4,useAMS,usenatbib,usegraphicx]{mn2e} 
\usepackage{amsmath} 
\usepackage{amssymb} 
\usepackage {graphicx}
\usepackage[dvips]{epsfig}
\usepackage{epsfig}  
\usepackage{color}
\usepackage[normalem]{ulem}
\usepackage{hyperref}
\usepackage{caption}
\usepackage{float}
\restylefloat{table}
\usepackage{array}
\usepackage{booktabs}
\def\be{\begin{equation}}
\def\ee{\end{equation}}
\def\ba{\begin{eqnarray}}
\def\ea{\end{eqnarray}}

\definecolor{red}{rgb}{1,0.0,0.0}

\definecolor{darkgreen}{rgb}{0.0,0.5,0.0}

\newcommand{\beq}{\begin{eqnarray}}  
\newcommand{\eeq}{\end{eqnarray}}  
 
\newcommand{\apj}{ApJ}  
\newcommand{\apjs}{ApJS}  
\newcommand{\apjl}{ApJL}  
\newcommand{\aj}{AJ}  
\newcommand{\mnras}{MNRAS}  
  
\newcommand{\aap}{A\&A}

\newcommand{\nat}{Nature}

\newcommand{\ly}{{\ifmmode{{\rm Ly}\alpha}\else{Ly$\alpha$}\fi}}
\newcommand{\hMpc}{{\ifmmode{h^{-1}{\rm Mpc}}\else{$h^{-1}$Mpc}\fi}}  
\newcommand{\hGpc}{{\ifmmode{h^{-1}{\rm Gpc}}\else{$h^{-1}$Gpc}\fi}}  
\newcommand{\hmpc}{{\ifmmode{h^{-1}{\rm Mpc}}\else{$h^{-1}$Mpc}\fi}}  
\newcommand{\hkpc}{{\ifmmode{h^{-1}{\rm kpc}}\else{$h^{-1}$kpc}\fi}}  
\newcommand{\hMsun}{{\ifmmode{h^{-1}{\rm {M_{\odot}}}}\else{$h^{-1}{\rm{M_{\odot}}}$}\fi}}  
\newcommand{\hmsun}{{\ifmmode{h^{-1}{\rm {M_{\odot}}}}\else{$h^{-1}{\rm{M_{\odot}}}$}\fi}}  
\newcommand{\Msun}{{\ifmmode{{\rm {M_{\odot}}}}\else{${\rm{M_{\odot}}}$}\fi}}  
\newcommand{\msun}{{\ifmmode{{\rm {M_{\odot}}}}\else{${\rm{M_{\odot}}}$}\fi}}

\newcommand{\rand}{{\ifmmode{{\mathcal{R}}}\else{${\mathcal{R}}$ }\fi}}  
\newcommand{\pr}[1]{ \left( #1 \right) }

\newcommand{\eq}[2]{\begin{equation} \label{eq:#1} #2 \end{equation}}

\newcommand{\reff}{{\ifmmode{r_{\mbox{\tiny eff}}}\else{$r_{\mbox{\tiny eff}}$}\fi}}
\begin{document}

%=========================================================================
%		FRONT MATTER
%=========================================================================
\title{Tensor anisotropy as a tracer of cosmic voids}
\author[S. Bustamante and J.E. Forero-Romero]{
\parbox[t]{\textwidth}{\raggedright 
  Sebastian Bustamante \thanks{sebastian.bustamante@udea.edu.co}$^{1}$,
  Jaime E. Forero-Romero \thanks{je.forero@uniandes.edu.co}$^{2}$ 
}
\vspace*{6pt}\\
$^1$Instituto de F\'{\i}sica - FCEN, Universidad de Antioquia, Calle
67 No. 53-108, Medell\'{\i}n, Colombia\\ 
$^2$Departamento de F\'{i}sica, Universidad de los Andes, Cra. 1
No. 18A-10, Edificio Ip, Bogot\'a, Colombia
}

\maketitle

\begin{abstract}
We present a new method to find voids in cosmological simulations
based on the tidal and the velocity shear tensors
definitions of the cosmic web. 
We use the fractional anisotropy (FA) computed from the eigenvalues
of each web scheme as a void tracer.
We identify voids using a watershed transform based on the local
minima of the FA field without making any assumption on the shape or
structure of the voids.  
We test the method on the Bolshoi simulation and report on the
abundance and radial averaged profiles for the density, velocity and
fractional anisotropy.
We find that voids in the velocity shear web are smaller than voids in
the tidal web, with a particular overabundance of very small voids in
the inner region of filaments/sheets.
We classify voids as subcompensated/overcompansated depending on the
absence/presence of an overdense matter ridge in their density
profile, finding that close to $65\%$ and $35\%$ of the total
population are classified into each category, respectively.
Finally, we find evidence for the existence of universal profiles from
the radially averaged profiles for density, velocity and fractional
anisotropy. 
This requires that the radial coordinate is normalized to the
effective radius of each void. 
Put together, all these results show that the FA is a reliable tracer
for voids, which can be used in complementarity to other existing
methods and tracers.
\end{abstract}


%=========================================================================
%		PAPER CONTENT
%=========================================================================

%*************************************************************************
\section{Introduction}
\label{sec:introduction}
%*************************************************************************

Cosmic voids are regarded as one of the most striking features of the
Universe on its larger scales ever since they were found in the first
galaxy surveys \citep{Chincarini75, Gregory78, Einasto80M, Einasto80N,
  Kirshner81, Zeldovich82,Kirshner87}.  
However, due to the large volume extension of void regions ($\sim
5-10\ \mbox{Mpc}  h^{-1}$), statistically meaningful catalogues of
voids \citep{Pan10,  Sutter12b, Nadathur14} have only become available
through modern galaxy surveys such as the two-degree field Galaxy
Redshift Survey (2dF) \citep{ Colless01, Colless03} and the Sloan
Digital Sky Survey (SDSS)\citep{York00, Abazajian03}.
These observational breakthroughs generated a great interest in the last
decade to study voids \citep{Hoyle04, Croton04, Padilla05, Rojas05,
  Ceccarelli06, Patiri06a, Tikhonov06, Patiri06b,Tikhonov07,
  BendaBeckmann08, Foster09, Ceccarelli13, Paz13, Sutter14a}.

On the theoretical side, the basic framework that explains
the origin of voids was established in the seminal work of
\citet{Zeldovich70} and refined in the following decades.  
The first detailed theoretical models describing formation, dynamics
and properties of  voids \citep{Hoffman82, Icke84, Bertschinger85,
  Blumenthal92} were  complemented and extended by numerical studies
\citep{Martel90, Regos91, Weygaert93, Dubinski93, Bond96}. 
Currently, the most popular approach to study voids relies on N-body
simulations. For an extensive compilation of previous  
numerical works we refer the reader to \citet{Colberg08}.

The relevance of voids to cosmological studies can be summarized in
three aspects \citep{Platen07}. 
Firstly, voids are a key ingredient of the Cosmic Web. 
They dominate the volume distribution at large scales and
compensate overdense structures in the total matter
budget. 
Secondly, voids provide a valuable resource to estimate  
cosmological parameters as their structure and dynamics are sensitive
to them. 
Finally, they are a largely pristine environment to test galaxy
evolution.

Although visual recognition of voids in galaxy surveys and simulations
is possible in most cases, we need a clear algorithmic identification
procedure to make statistical studies.
Nevertheless, the community has not reached yet an unambiguous
definition of cosmic voids.
There are many different void finding techniques in the literature
(for a detailed comparison of different schemes,  see the publication
on the results of the Void Finder Comparison Project \citet{Colberg08}).  
In spite of the diversity of existing schemes, they can be roughly
classified into two types: point-based and field-based. 
There are geometric schemes based on point distributions (either real or
redshift space)  f galaxies in surveys or dark matter halos in
simulations \citep{Kauffmann91, Muller00,    Gottlober03, Hoyle04,
  Brunino07,  Foster09, Micheletti14, Sutter14}. 
While other schemes are based on the smooth and continuous matter density
field either from simulations or from reconstruction procedures on
surveys \citep{Plionis02, Colberg05,  Shandarin06, Platen07,
  Neyrinck08, MunozCuartas11, Neyrinck13, Ricciardelli2013}. 
Our work follows the tradition of the second kind of schemes.

Here we introduce a new algorithm to define voids over the continuous
matter density or velocity distribution defined on a fixed and
homogeneous spatial grid.
The algorithm uses the results from two tensorial schemes used to
classify the cosmic web.
The first (the T-Web) is based on the Hessian of the gravitational potential or
tidal tensor \citep{Hahn07, Forero09}. 
The second (the V-web) is based on the velocity shear tensor
\citep{Hoffman12}. 
Our procedure allows for a description of a void internal structure
beyond a simple definition of a void as an underdense region in the
large-scale matter distribution. 
The tidal and the shear tensors
encode more information than the density/velocity fields as they trace
the collapsing or expanding nature of the matter field, which defines
the dynamics of the Cosmic Web.

The tracer that we use to define the voids is the fractional
anisotropy (FA) computed from the set of eigenvalues of the tensor
under consideration. 
The FA was initially introduced by \citet{Basser95} to quantify the
anisotropy degree of the diffusivity of water molecules through
cerebral tissue in nuclear magnetic resonance imaging, thereby allowing to 
detect structures that restrict the otherwise isotropic Brownian movement 
of water molecules. Taking into account that this process is identical to 
the velocity shear in the large-scale matter distribution, 
\citet{Libeskind13} introduced the concept of the FA in the context of 
Cosmic Web classification schemes.

%!!!!!!!!!!!!!!!!!!!!!!!!!!!!!!!!!!!!!!!!!!!!!!!!!!!!!!!!!!!!!!!!!!!!!!!!!

In the next sections (\S \ref{sec:algorithms_cosmic_web} and \S
\ref{sec:bulk_voids}) we establish the FA as a void tracer and then
proceed to identify individual voids as basins of FA local minima.  At
this point we implement a \textit{watershed transform algorithm}
\citep{Beucher79,Beucher93} which has been used to define voids as
catching basins of local minima of the density field \citep{Platen07,Neyrinck08}.
Finally we find and charachterize voids in a N-body (\S
\ref{sec:simulations} and \S \ref{sec:results}) using density, velocity and
fractional anisotropy profiles.  We use these results to comment on
the qualities of our algorithm (\S \ref{sec:conclusions}).

%*************************************************************************
\section{Algorithms to find the Cosmic Web}
\label{sec:algorithms_cosmic_web}
%*************************************************************************

%.........................................................................
%FIGURE 1: FA and vissual impression
\begin{figure*}
  \includegraphics[trim = 16mm 8mm 5mm 12mm, clip, keepaspectratio=true,
  width=0.73\textheight]{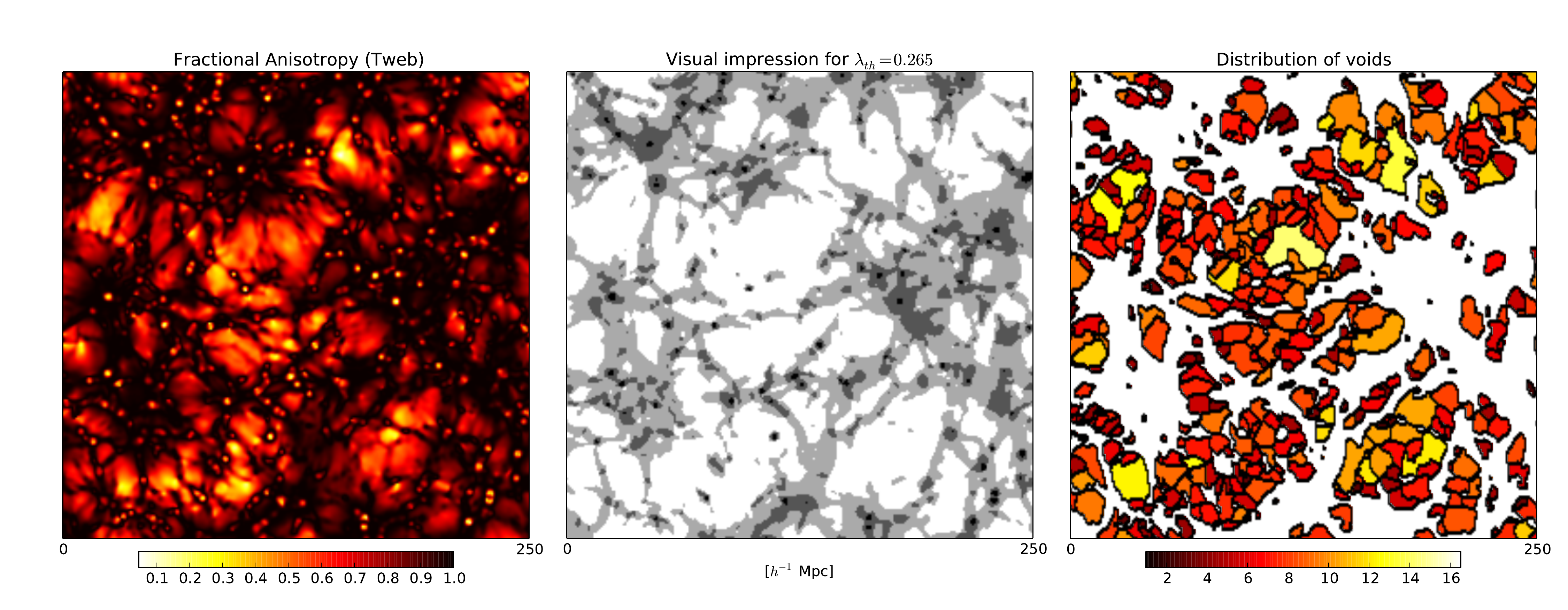}
  \includegraphics[trim = 16mm 8mm 5mm 12mm, clip, keepaspectratio=true,
  width=0.73\textheight]{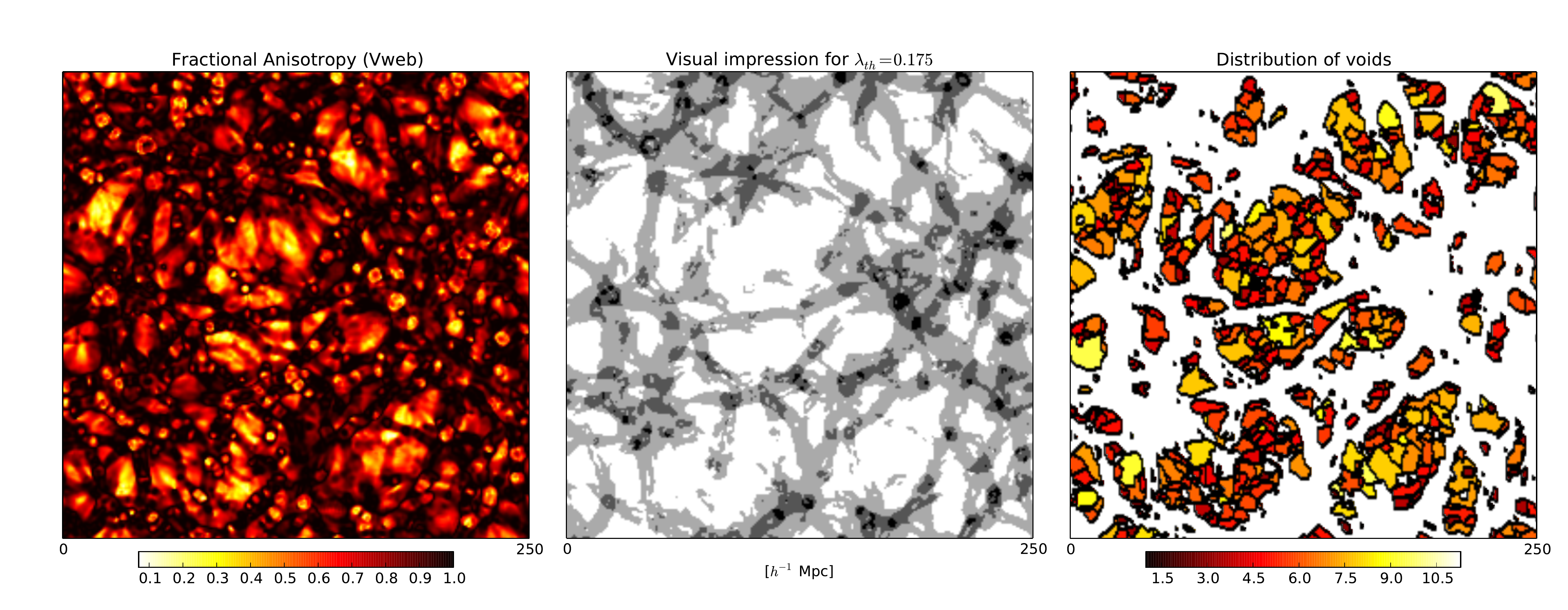}
  \caption{Different fields in a slice of
    $1\hMpc$ width in the Bolshoi simulation. Top/bottom panels correspond
    to the T-Web/V-Web.      
    Fractional Anisotropy field (left); cosmic web classification
    (middle) for a given value of the threshold eigenvalue
    $\lambda_{\rm th}$ where voids are white, sheets are light gray,
    filaments are dark gray and knots are black; individual voids (right) found by 
    the a watershed algorithm on the Fractional Anisotropy field. Colours
    in the right panel correspond with the effective radius of each void. }
  \label{fig:FA_field}
\end{figure*}
%.........................................................................

Our new void finding method is based on two existing cosmic web
classification schemes that work on cosmological N-body simulations.
Both schemes depend on the construction of tensors based on the
Hessian of the potential (T-Web scheme) and the shear of the velocity
(V-Web scheme). 
These algorithms have been used to develop other
kind of studies such as the alignment of the shape, spin and peculiar
velocity of dark matter halos with the cosmic web
\citep{Libeskind13,Forero2014}. 
Here we summarize the most relevant aspects of each scheme. 
We refer the reader to the papers of \cite{Forero09} (T-Web) and
\cite{Hoffman12} (V-Web) for detailed descriptions.

%-------------------------------------------------------------------------
\subsection{The tidal web (T-Web)}
\label{subsec:Tweb}
%-------------------------------------------------------------------------

This scheme was initially proposed by \citet{Hahn07}  to find the
Cosmic Web based on the tidal tensor. 
The tidal tensor allows a classification in terms of the orbital
dynamics of the matter field.
This approach extends to second-order the equations of motion around 
local minima of the gravitational potential. 
The second-order term corresponds to the tidal tensor, which is
defined as the Hessian matrix of the normalized gravitational
potential

%.........................................................................
%Tidal Tensor
\eq{V_web}
{	T_{\alpha\beta} = \frac{\partial^2\phi}{\partial x_{\alpha}\partial x_{\beta}},	}
%.........................................................................
where the physical gravitational potential has been rescaled by a
factor of   $4\pi G\bar{\rho}$ in such a way that $\phi$ satisfies the following 
Poisson equation

%.........................................................................
%Poisson
\eq{Poisson}
{	\nabla^2\phi = \delta,	}
%.........................................................................
with $\bar{\rho}$ the average density in the Universe, $G$ the 
gravitational constant and $\delta$ the dimensionless matter
overdensity.

Since the tidal tensor can be represented by a real and  symmetric
$3\times 3$ matrix, it is always possible to diagonalize  
it and obtain three real eigenvalues $\lambda_{1}\geq\lambda_{2}\geq
\lambda_3$ with its corresponding eigenvectors ${\bf u}_{1}$, ${\bf u}_{2}$,
${\bf u}_{3}$. 
The eigenvalues are indicators of the local  orbital stability
in each direction ${\bf u}_i$. 
The sign of the eigenvalues can be used to classify the Cosmic Web.
The number of positive (stable) or negative (unstable) eigenvalues allows 
to label a location into one of the next four types of environment: 
voids (3 negative eigenvalues), sheets (2), filaments (1) and knots (0).

A modification to this scheme was introduced by \citet{Forero09}
by means of a relaxation of the stability criterion. 
The relative strength  of each eigenvalue is no longer defined by the
sign, but instead by a threshold value $\lambda_{\rm th}$ that can be
tuned in such a way that the visual impression of the web-like matter
distribution is reproduced. 
 
%-------------------------------------------------------------------------
\subsection{The velocity web (V-Web)}
\label{subsec:Vweb}
%-------------------------------------------------------------------------

The V-web scheme for environment finding introduced by
\cite{Hoffman12} is based on the local velocity shear tensor
calculated from the smoothed dark matter  velocity field in the
simulation. 
This tensor is given by the  following expression

%.........................................................................
%V-Web Definition
\eq{V_web}
{	\Sigma_{\alpha\beta} = -\frac{1}{2H_0}\pr{\frac{\partial v_{\alpha}}
{\partial x_{\beta}}+\frac{\partial v_{\beta}}{\partial x_{\alpha}}},}
%.........................................................................
where $v_{\alpha}$ and $x_{\alpha}$ represent the $\alpha$ component of 
the comoving velocity and position, respectively. Like the tidal tensor, 
$\Sigma_{\alpha\beta}$ can be represented by a $3\times 3$ symmetric 
matrix with real values, making it possible to find three real
eigenvalues and its corresponding eigenvectors.

In this case we also use the relative strength of the three eigenvalues with 
respect to a threshold value $\lambda_{th}$ to classify the cosmic web
in the four web types already mentioned.

\begin{figure*}
\centering
  \includegraphics[trim = 1mm 0mm 5mm 10mm, keepaspectratio=true,
  width=0.33\textwidth]{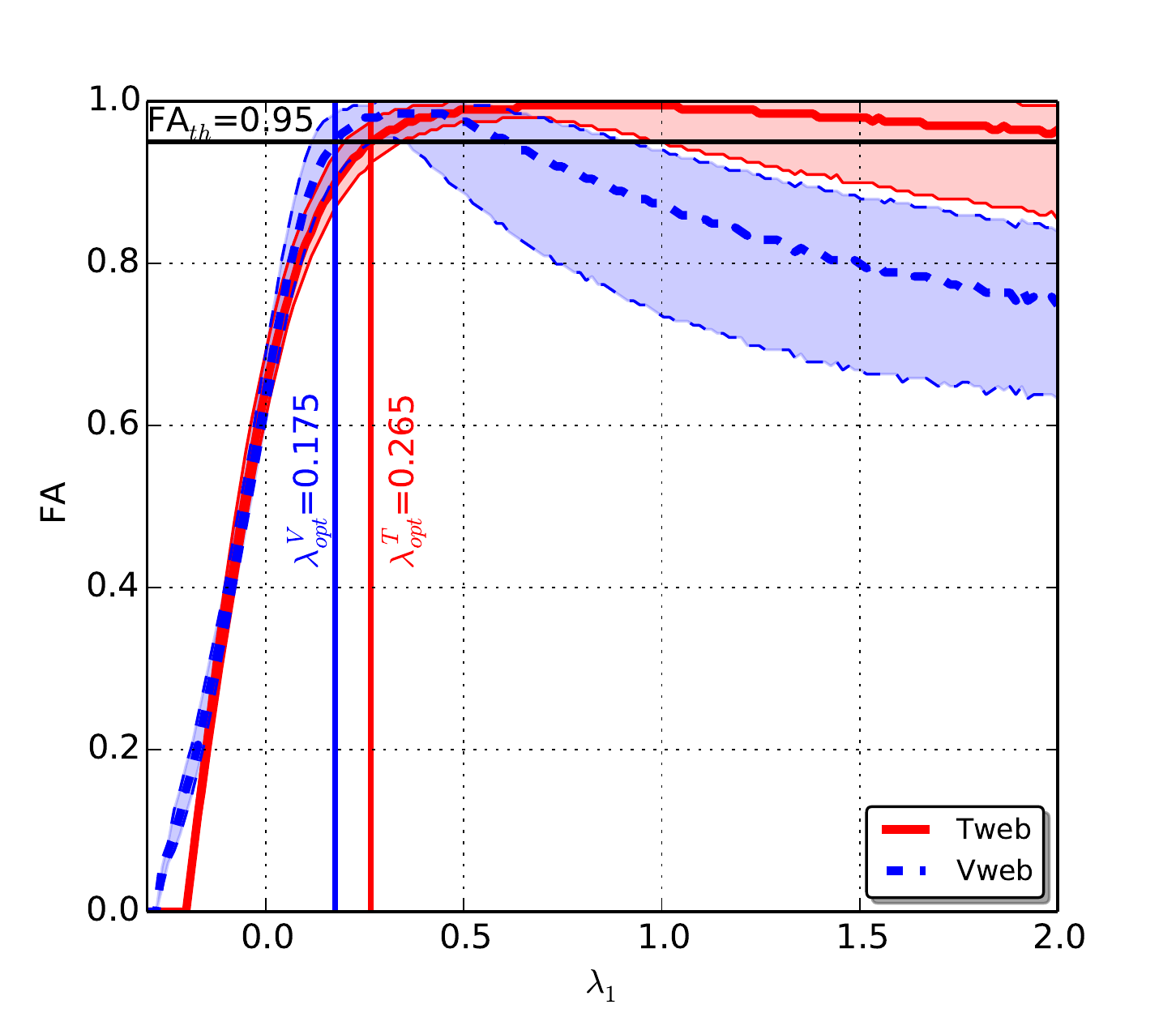}  
  \includegraphics[trim = 1mm 0mm 5mm 10mm, keepaspectratio=true,
  width=0.33\textwidth]{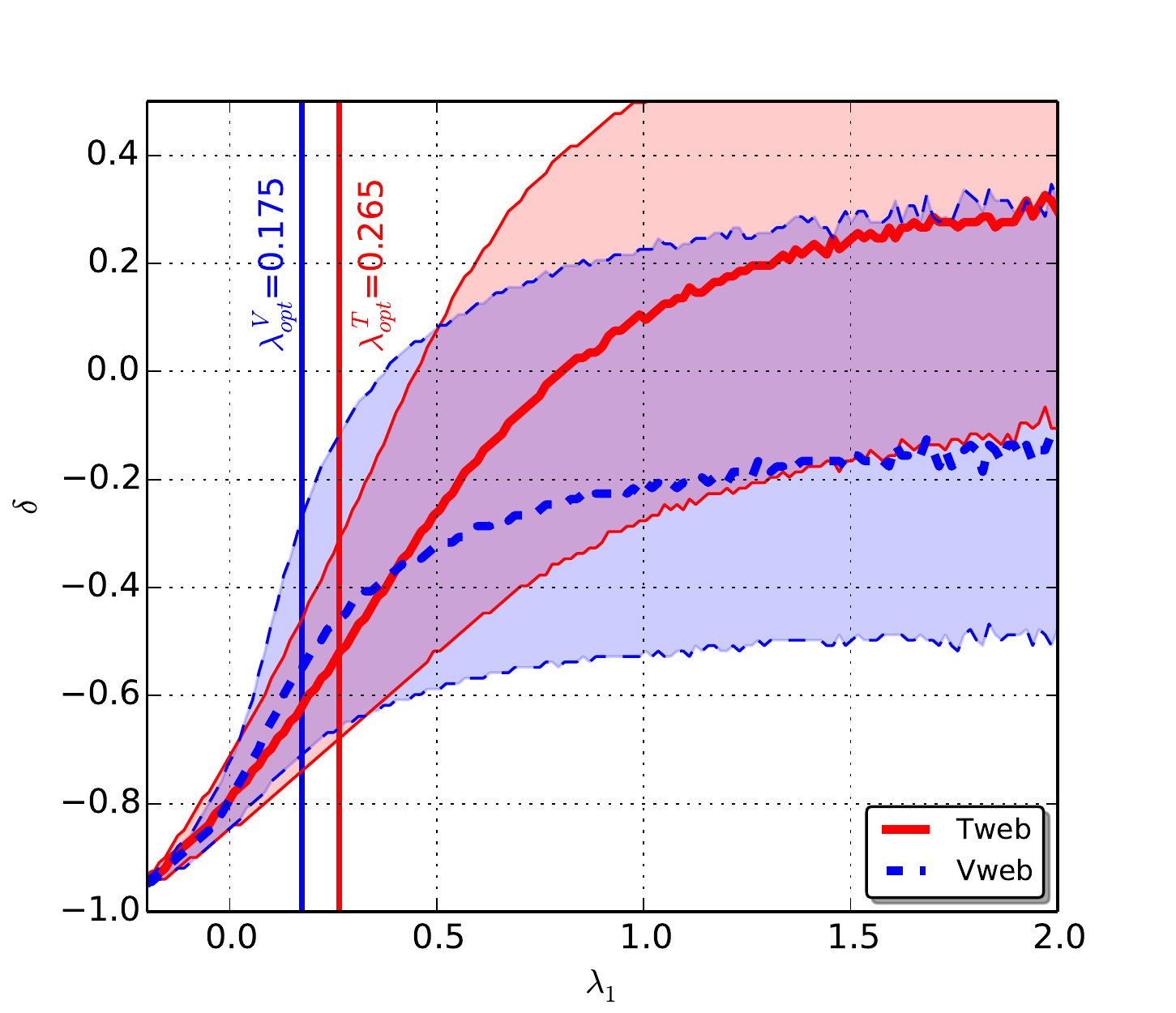}
  \includegraphics[trim = 1mm 0mm 5mm 10mm, keepaspectratio=true,
  width=0.33\textwidth]{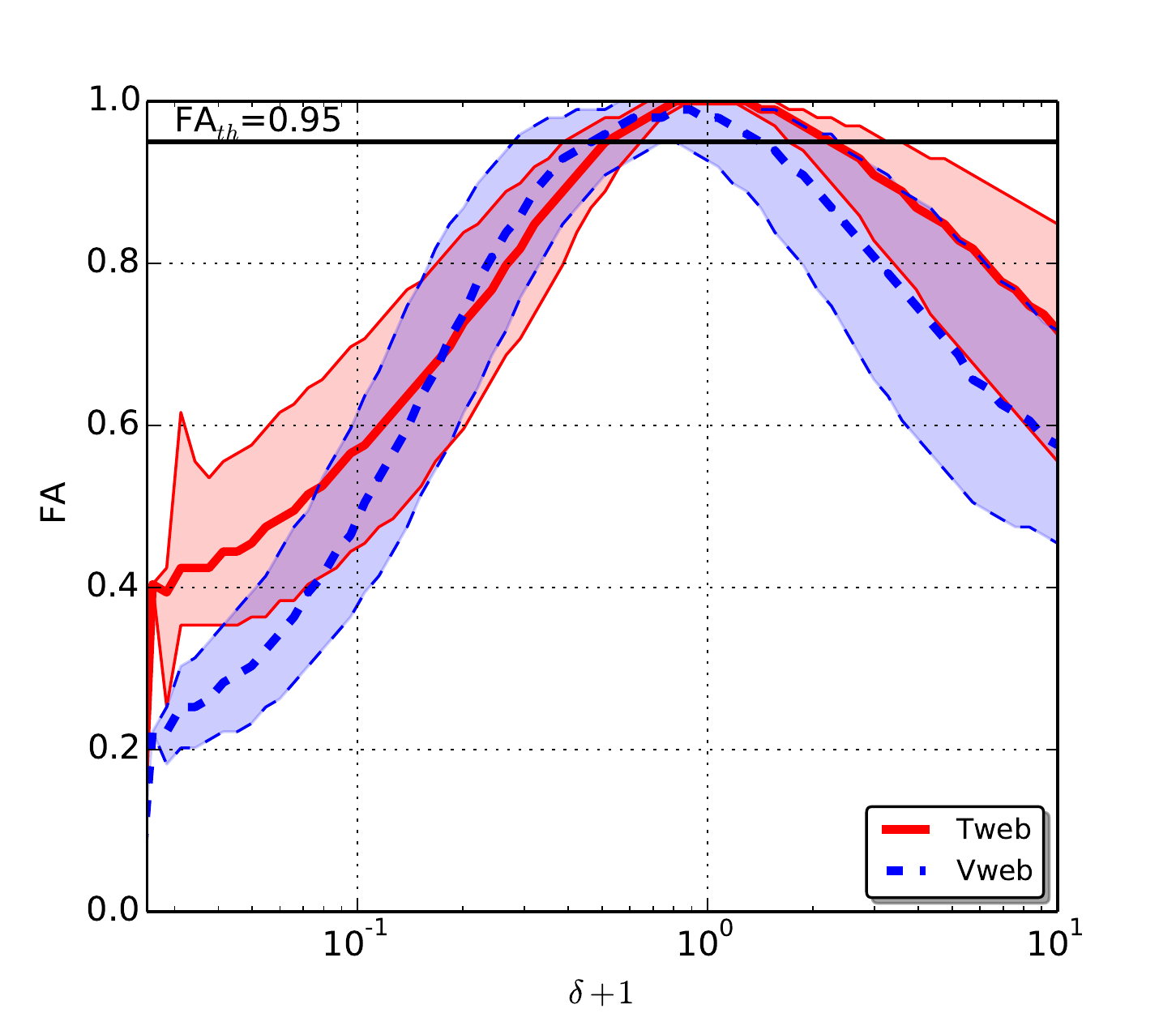}
  \captionof{figure}{
    Fractional Anisotropy  as a function of
    the overdensity (left) and the eigenvalue $\lambda_1$ (middle), and
    density contrast (right) as a function of the
    eigenvalue $\lambda_1$ for each web scheme (T-Web,  continuous
    lines. V-Web, dashed lines).     
    Fractional Anisotropy  as a function of $\lambda_1$ (left),
    overdensity as a function of $\lambda_1$ (middle) and Fractional
    Anisotropy as a function of the overdensity (right).
    Thick central lines are the median
    and filled regions include $50\%$ of the cells in each bin. 
    The left panel shows the tight correlation between FA and
    $\lambda_1$ in voids.
    In the middle and right panel we show the equivalent eigenvalue
    threshold for the T-Web and V-web corresponding to FA=$0.95$. 
    The right panel confirms that underdense regions can be
    identified with Fractional Anisotropy values less than $0.95$.
    For this reason we define voids to be composed by cells with
    Fractional Anisotropy equal or less than this value.
    The threshold $\lambda_{\rm th}$ on the eigenvalues does not
    influence our results. 
    We show it as a guide to readers familiar with the
    order of magnitude expectation for this quantity.}
  \label{fig:L1_correlations}

\end{figure*}

Usually, the threshold $\lambda_{th}$ is a free parameter that is
tuned to reproduce the visual appearance of the comic web. 
However, in this paper this threshold does not play any role in our
computations.
In Figure \ref{fig:L1_correlations} we offer a threshold estimate
(from the FA=$0.95$ as a function of $\lambda_1$, this choice is
explained in the next section) only as a guide to readers familiar
with its meaning.

%-------------------------------------------------------------------------
%FIGURE 2: Distributions of FA and density regarding the Lambda_1 eigenvalue
%.........................................................................

%-------------------------------------------------------------------------
%FIGURE 3: Histograms for each environment regarding the FA value
\begin{figure}
\centering
  \includegraphics[trim = 1mm 0mm 5mm 10mm, clip, keepaspectratio=true,
  width=0.35\textheight]{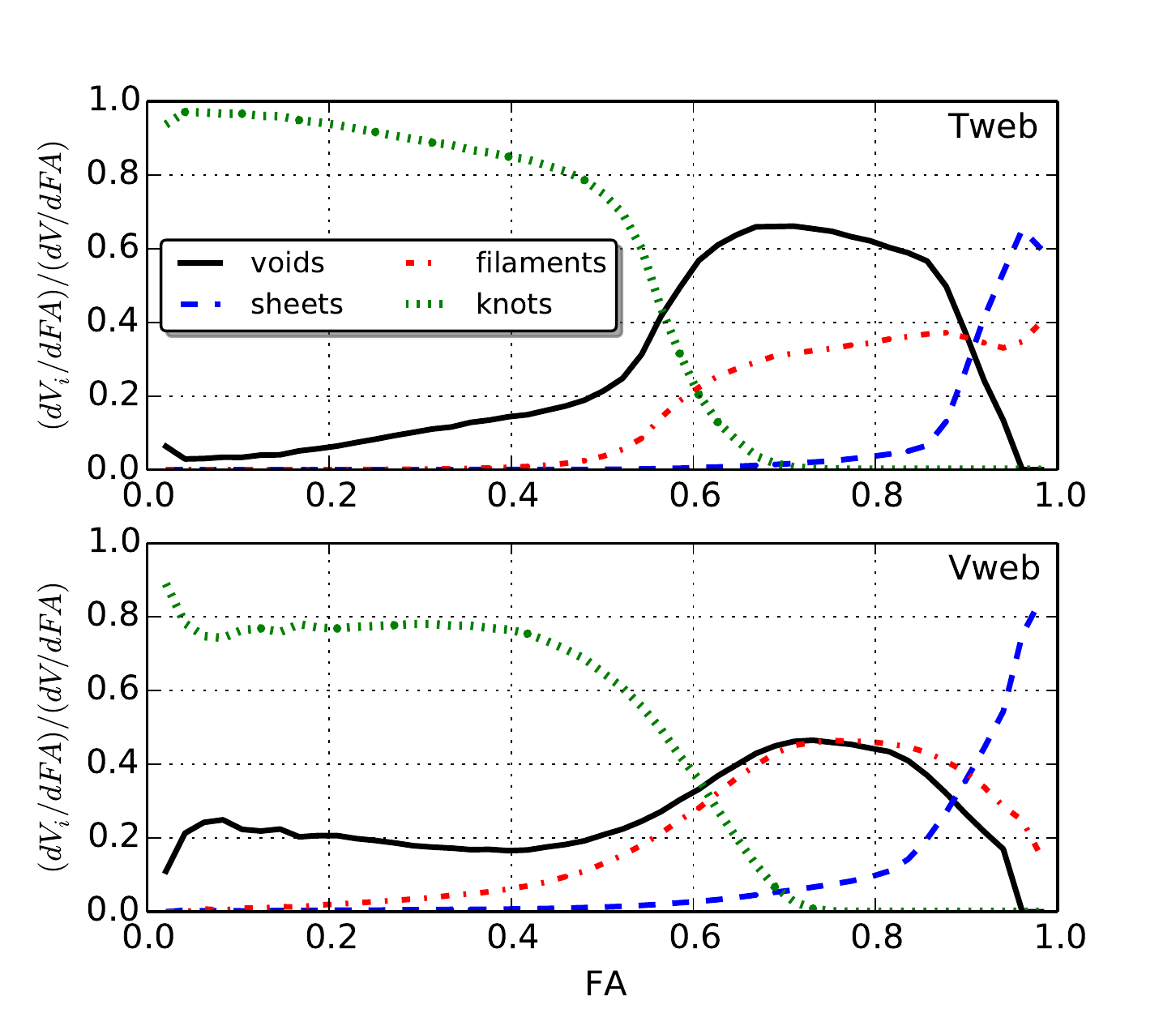}
   \captionof{figure}{\small Volume of each environment per FA
   ($dV_i/dFA,\ \ i=v,s,f,k$) normalized by the total volume per FA
   ($dV/dFA$). 
  Top/bottom corresponds to the T-Web/V-Web. For a given value of the FA, 
  each line represents the volume contribution of the respective 
  environment to the total volume with the same FA value. Above FA=0.95,
  only sheets and filaments contribute to all the volume of the simulation
  within that range of FA.}
  \label{fig:FA_environment}
\end{figure}
%.........................................................................

%*************************************************************************
\section{A new void finding technique}
\label{sec:bulk_voids}
%*************************************************************************

%-------------------------------------------------------------------------
\subsection{The fractional anisotropy}
\label{subsec:FA_voids}
%-------------------------------------------------------------------------

% CORRECTION !!!!!!!!!!!!!!!!!!!!!!!!!!!!!!!!!!!!!!!!!!!!!!!!!!!!!!!!!!!!!

The fractional anisotropy (FA), as developed by \citet{Basser95}, 
was conceived to quantify the anisotropy of the diffusivity of water
molecules through cerebral tissue in nuclear magnetic resonance imaging. 
In this manner, barriers such as microtubules and cell membranes, that 
restrict the otherwise isotropic Brownian motion of water molecules, can 
be detected. In a cosmological context, the problem of finding structures 
is very similar, where filaments and sheets restrict the otherwise 
isotropic collapsing/expanding dynamics of the matter. Taking this into 
account, we present here the FA, much in the same way as \citet{Libeskind13}, 
to use it as a tracer of cosmic voids.

%!!!!!!!!!!!!!!!!!!!!!!!!!!!!!!!!!!!!!!!!!!!!!!!!!!!!!!!!!!!!!!!!!!!!!!!!!

The FA is defined as follows
%.........................................................................
%Fractional anisotropy
\eq{fractional_anisotropy}
{{\mathrm{FA}} = \frac{1}{\sqrt{3}}\sqrt{ \frac{ (\lambda_1 - \lambda_3)^2 + 
(\lambda_2 - \lambda_3)^2 + (\lambda_1 - \lambda_2)^2}{ \lambda_1^2 + 
\lambda_2^2 + \lambda_3^2} },}
%.........................................................................
where the eigenvalues can be taken from either the T-Web or the V-Web 
(FA-T-Web and FA-V-Web respectively). Such as it is defined, FA$=0$ 
corresponds to an isotropic distribution ($\lambda_1=\lambda_2=\lambda_3$) 
and FA$=1$ to a highly anisotropic distribution.

In the left and middle panels of Fig. \ref{fig:FA_field} we show the
FA field and web classification for both  web schemes over a slice of
an N-body simulation (described in Section \ref{sec:simulations}). 
Comparing these two panels we see that voids and knots (white
and black in the middle panel of Fig. 1) display low FA values at
their  centres, becoming gradually more anisotropic at outer regions.  
On the other hand the filamentary structure (grey in the middle panel
of Fig. 1) is traced by middle to high FA values.  
This characteristic is the  key to use the FA as a tracer of cosmic
voids.

%-------------------------------------------------------------------------
\subsection{Fractional anisotropy as a void tracer}
\label{subsec:web_voids}
%-------------------------------------------------------------------------

Voids are regions where $\lambda_3\leq\lambda_2\leq
\lambda_1\leq\lambda_{\rm th}$. 
This implies that a void is completely fixed by the relative strength
of the $\lambda_1$ eigenvalue with respect to the threshold.   
As we increase/decrease the threshold value $\lambda_{\rm th}$, voids
increase/decrease progressively through contours of
increasing/decreasing $\lambda_1$.  
Voids are thus characterized by low values of both FA and
$\lambda_1$.

In Fig. \ref{fig:L1_correlations}  we show that the these two values
are correlated. 
The left panel shows the correlation between the FA and $\lambda_1$
and the middle panel between the overdensity $\delta$ and $\lambda_1$.
The right panel shows the correlation between $\delta$ and the
FA. 
This shows that the  density has a larger scatter at fixed
$\lambda_1$ than the FA, yet large and  small values of the density
remain still associated to low values of the FA.

At this point, the eigenvalue $\lambda_1$ and the FA appear to be
equally potential candidates for tracing cosmic voids, however, the 
eigenvalue $\lambda_1$ presents, along with the overdensity, an undesirable 
feature for a void tracer. 
As we move out from inner parts of voids to outer parts, $\lambda_1$
and $\delta$ increase monotonically making unclear where to set the
void boundary, while for low density regions ($\delta<1$) the FA traces
the boundary where anisotropic collapsing regions begin to 
dominate, i.e. at the ridge reached around $\delta = 0$ as shown in the 
right panel of Fig. \ref{fig:L1_correlations}.
According to Fig. \ref{fig:FA_environment} these anisotropic areas are
mainly associated to sheets and, to a lesser extent, filaments.

From Figs. \ref{fig:FA_field} and \ref{fig:L1_correlations} we
conclude that the FA is a good tracer of voids as it is almost perfectly
correlated with low values of $\lambda_1$. 
We propose that voids should be composed completely by regions of
FA$<0.95$.
If we increase the values of $\lambda_1$ from its minimum until it
we reach FA$=0.95$ in \ref{fig:L1_correlations} we find that this
correspond to critical values of $\lambda_{1}^T = 0.265$ and
$\lambda_{1}^V = 0.175$ for the T-Web and V-Web, respectively.
This means that setting $\lambda_{th}$ to either
$\lambda_{1}^T$/$\lambda_{1}^{V}$ automatically produces voids with
all the cells FA$<0.95$.   
The middle panels in Fig. \ref{fig:FA_field} show the web
classification for this choice of $\lambda_{th}$, demonstrating that
this FA level is a reasonable choice to define voids.

%-------------------------------------------------------------------------
\subsection{Defining voids with a watershed algorithm}
\label{subsec:watershed}
%-------------------------------------------------------------------------

The previous Section shows that FA is a good void tracer, but how
should we actually define the boundary of individual voids?
For this purpose, we use the \textit{watershed transform algorithm}
\citep{Beucher79,Beucher93} to identify a void as the basin of FA
local minima with boundaries of FA$=0.95$. 
The advantage of this definition is that it does not require any
assumption on the shape and/or morphology of the tentative voids. 

However, there are two main differences in our approach with respect to other
watershed implementations.
First, the watershed technique commonly uses the density field
instead of the FA field as we do here
\citep{Platen07,Neyrinck08}.   
Second, we estimate all relevant quantities on a Cartesian mesh of
fixed cell size, while other works use an adaptive Delaunay tessellation
\citep{Schaap00}.
Nevertheless, in our case the average number of particles per cell
  is large enough ($512$ particles/cell) ensuring that the lowest
  density regions are sampled with at least $\sim 10$ particles
  ($\delta \sim-0.98$), which seems to avoid spurious results from a
  noisy density field sampling.

The watershed algorithm also needs a threshold value to reduce
spurious features and prevent void hierarchization.  
If the density field is used, a typical threshold is
$\delta = -0.8$ \citep{Platen07},  which means that, given two neighbouring 
voids, if the mean value of the density across their common boundary is 
below that threshold, the two voids are merged.
In our case we have to find a corresponding FA value
to define this threshold.

In the right panel of Fig. \ref{fig:L1_correlations} we see how 
the FA correlates with the matter overdensity. 
A value of $\delta=-0.8$ corresponds to a value of FA$\sim 0.65$
regardless of the web-finding scheme. 
This is the value we choose to remove ridges between two watershed
voids. 
The right column in Fig. \ref{fig:FA_field} shows all the individual
voids that have been identified using the watershed algorithm on the
FA field.

%.........................................................................
%FIGURE 4: volume functions
\begin{figure*}
\centering

  \includegraphics[trim = 2mm 2mm 2mm 0mm, clip, keepaspectratio=true,
  width=0.35\textheight]{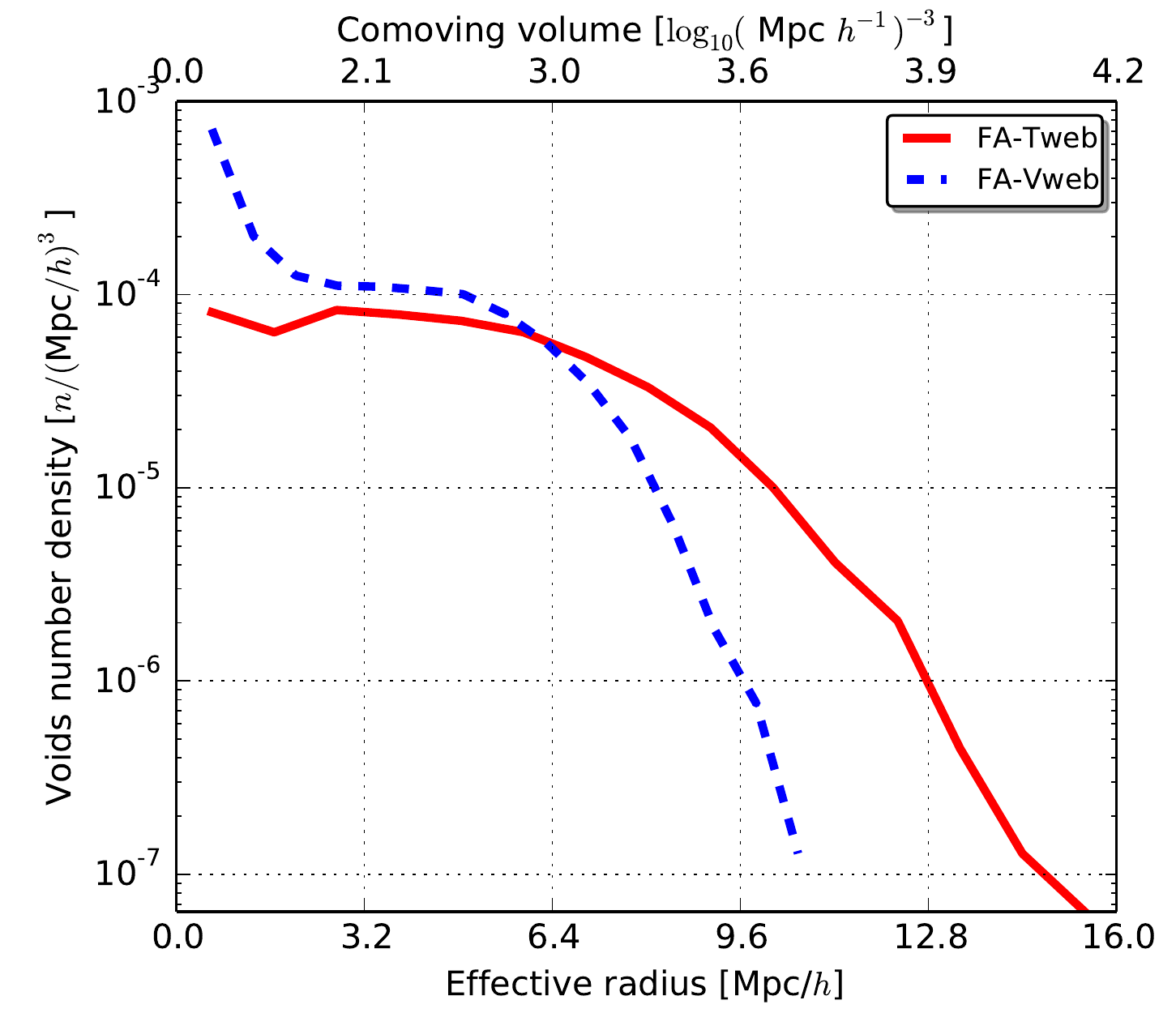}
  \includegraphics[trim = 2mm 2mm 2mm 0mm, clip, keepaspectratio=true,
  width=0.35\textheight]{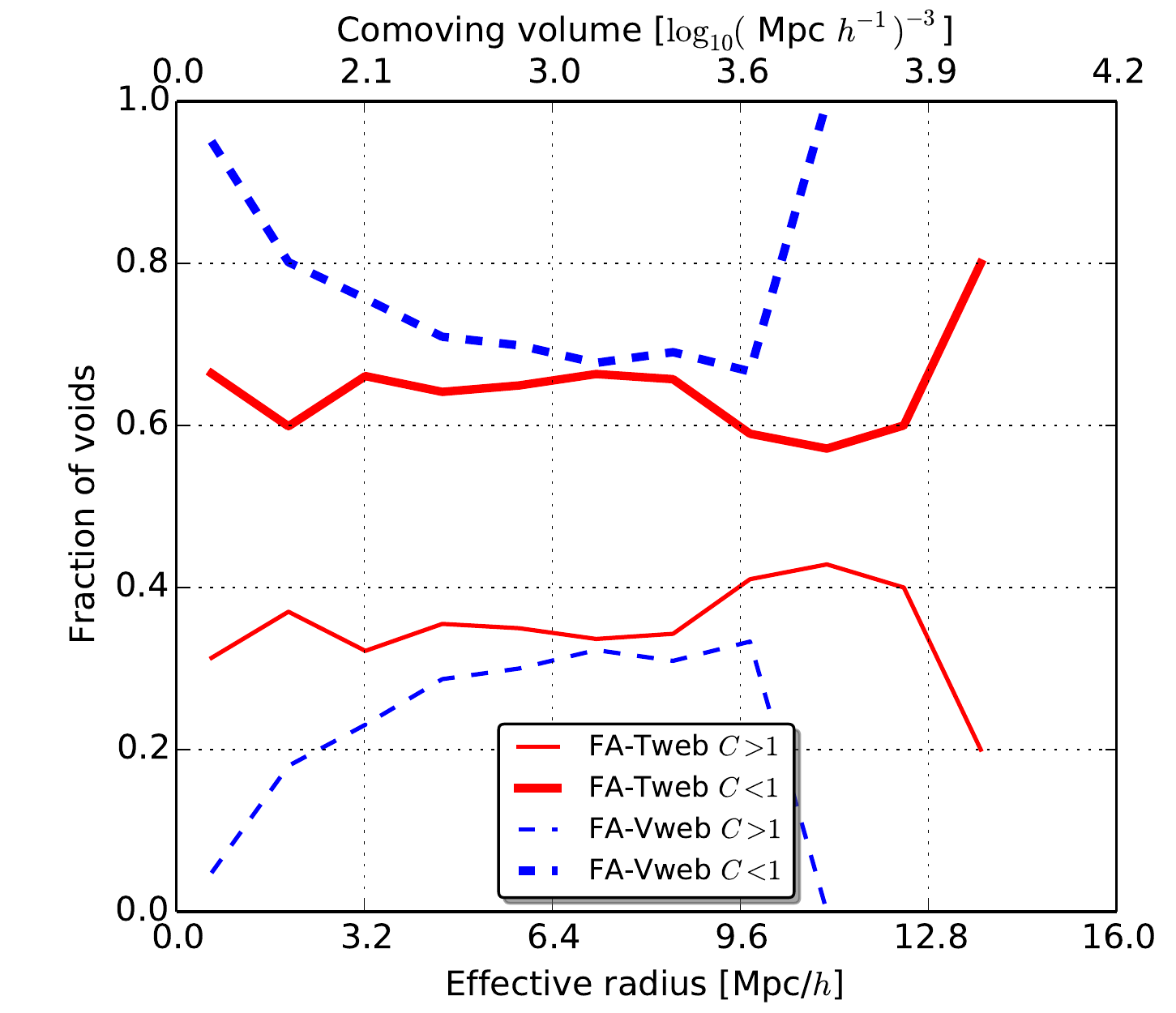}

  \captionof{figure}{\small Left panel, void size distribution. Right
    panel, fraction of overcompensated/subcompensated ($C>1/C<1$)voids as a
    function of the effective radius. The continuous (dashed)
    curves correspond to the T-Web (V-Web).
    The abundance for small ($\reff <2.0\hMpc$) overcompensated voids
    shoots up in the V-Web scheme.
    We provide in the body of the paper two reasons for this behaviour.}
  \label{fig:volume_function}

\end{figure*}
%.........................................................................

\section{Numerical Simulation}
\label{sec:simulations}

We use the Bolshoi simulation to test our void finding method. 
This simulation follows the non-linear evolution of a dark
matter density field on a cubic volume of  size $250$\hMpc\ sampled
with $2048^3$ particles. 
The cosmological parameters in the simulation are $\Omega_{\rm
  m}=0.27$,  $\Omega_{\Lambda}  =0.73$, $h=0.70$, $n=0.95$ and
$\sigma_{8}=0.82$ for the matter density,   cosmological 
constant, dimensionless Hubble parameter, spectral index of
primordial density perturbations and normalization for the power
spectrum, respectively. 
These values are consistent with the ninth year of data of the
Wilkinson  Microwave Anisotropy Probe (WMAP) \citep{Hinshaw2013}. 
For more detailed  technical information about the simulation, see
\citet{Klypin11}.

We use data for the cosmic web identification that is publicly
available through the MultiDark database
\url{http://www.multidark.org/MultiDark/} which is described in
\citet{Riebe11}. 
Here we briefly describe the process to obtain the data. 
For details see \cite{Forero09} (T-Web);
\cite{Hoffman12,Forero2014} (V-Web). 
This data is based on a  \textit{cloud-in-cell} (\texttt{CIC})
interpolation of the density and velocity fields of the simulation
onto a grid of $256^3$  cells, corresponding to a spatial resolution
of $0.98$\hMpc\ per cell side.    
These fields are smoothed with a gaussian filter with a width of 
$\sigma=0.98$\hMpc. 
The tidal and shear tensors and corresponding eigenvalues are computed
through finite-differences over the potential and velocity fields.

%*************************************************************************
\section{Results}
\label{sec:results}
%*************************************************************************

We limit our results to voids with effective radius larger than the
smoothing length of the density field, i.e. $\sim 1$\hMpc.  
For smaller interpolation/smoothing scales the shot noise in
  the low density regions starts to be noticeable.
With that choice we find a void volume filling fraction $54.88\%$ and
$47.06\%$ for the FA-Vweb and the FA-Tweb, respectively.  
For larger smoothing scales ($>5\hMpc$) the density and shear fields
  tend to be Gaussian and therefore follow closely the analytical
  environment abundance predictions for Gaussian random fields
  \citep{Forero09,Alonso2015}.  
In this paper we limit ourselves to a smoothing scale $\sim 1$ \hMpc.

In the following subsections we describe the results for the size
distribution and different radial profiles for all our samples.

%-------------------------------------------------------------------------
\subsection{The void size distribution}
\label{subsec:shape_voids}
%-------------------------------------------------------------------------

Void shapes exhibit a wide range of geometries.
To define their size we use its equivalent spherical radius or
effective radius, defined as $\reff = [3/(4\pi)V]^{1/3}$, with $V$ the 
total volume of the void computed from the individual grid cells
assigned to the void.   
In Fig. \ref{fig:volume_function} we show the void size
distributions for the T-Web and the V-Web.

We see that the void distribution for the T-Web is broadly consistent with the
expectations from a two-barrier problem  \citep{Sheth04}. 
The formation of large voids is limited by the \textit{void-in-void}
mechanism (first barrier) where large voids are constituted
hierarchically of smaller ones. 
In turn, the formation of small voids is damped by the
\textit{void-in-cloud} mechanism (second barrier),  where nearby
collapsing structures limit the abundance of small embedded voids.  

We also find that the V-Web scheme produces an over-abundance of small
voids compared to T-Web.
A large number of these small voids are embedded in overdense regions. 
They are already visible in the middle panel of Fig. \ref{fig:FA_field} as
tiny white bubbles located inside sheets.
The existence of these small voids can be explained by
dynamics of shell crossing in collapsing sheets as discussed in
\citet{Hoffman12}.    
The main argument is that as matter collides into a pancake-like
fashion crossing sheets encounter each other at the symmetry plane.
This effectively gives a positive divergence in the velocity
field, resulting in a void identification by the V-Web algorithm.

Comparing the abundance of large voids in the two web schemes, we find
that the largest voids in the V-Web have $\reff \sim 10$\hMpc,
while the T-Web scheme includes voids as large as $\reff \sim 15$\hMpc. 
This suggests that large voids in the T-Web scheme have a velocity
structure that splits them in the V-Web. 
This increased granularity in the velocity structure of voids is
evident in the left and right panels of Fig. \ref{fig:FA_field}.

%.........................................................................
%FIGURE 5: Density profile of voids for each defined scheme
\begin{figure*}
\centering
  \includegraphics[trim = 2mm 2mm 5mm 0mm, clip, keepaspectratio=true,
  width=0.35\textheight]{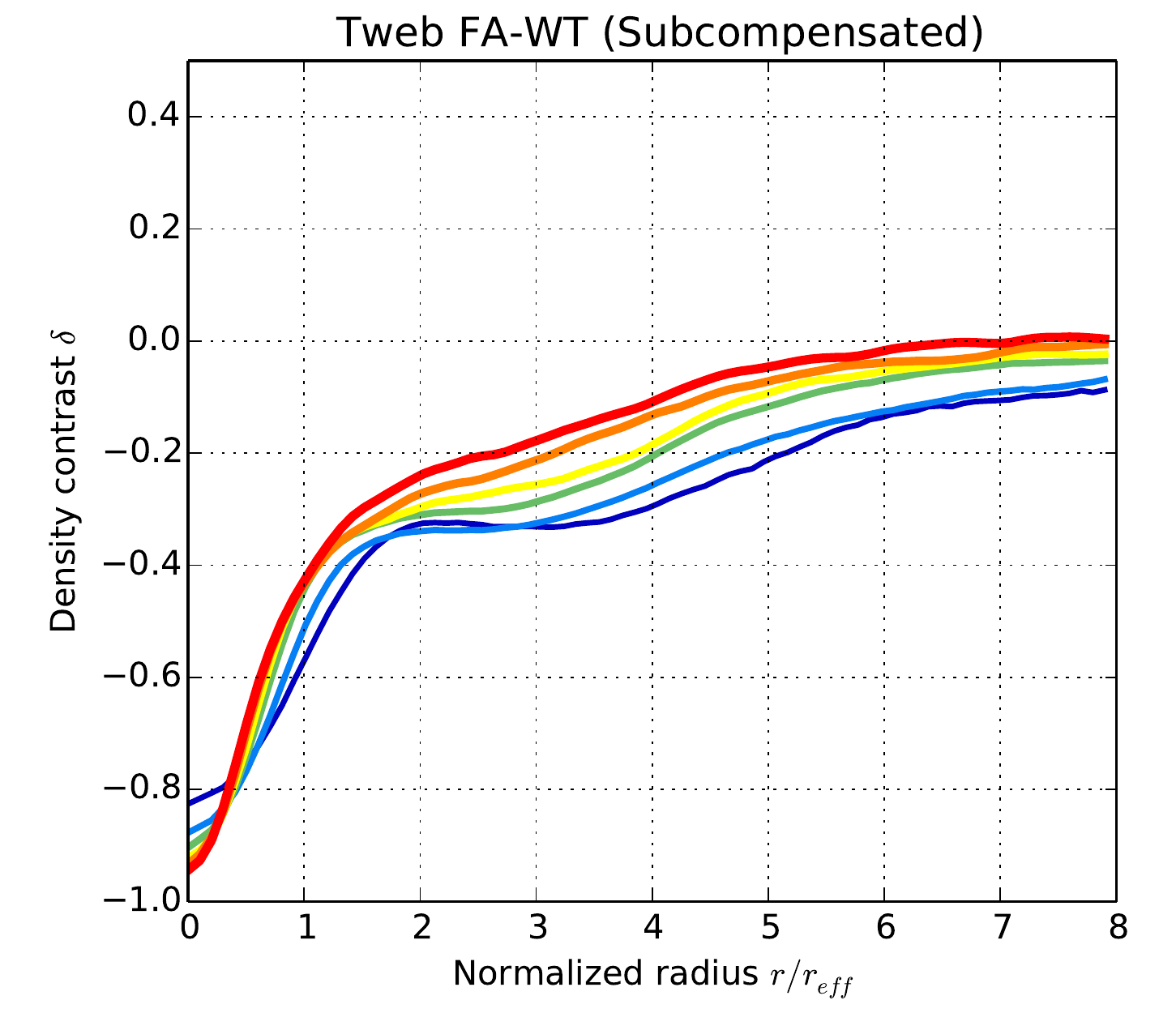}
  \includegraphics[trim = 2mm 2mm 5mm 0mm, clip, keepaspectratio=true,
  width=0.35\textheight]{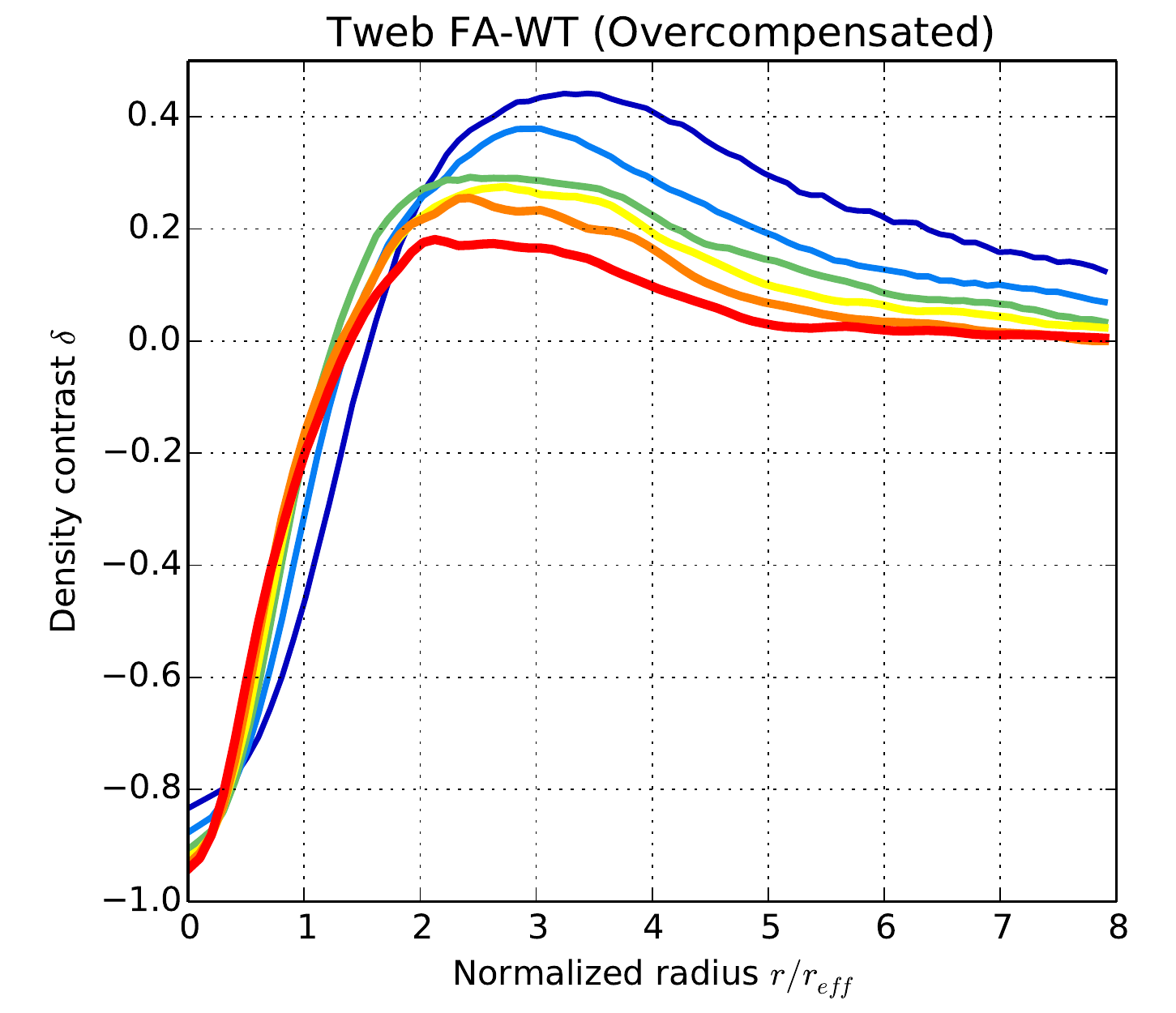}

  \includegraphics[trim = 2mm 2mm 5mm 0mm, clip, keepaspectratio=true,
  width=0.35\textheight]{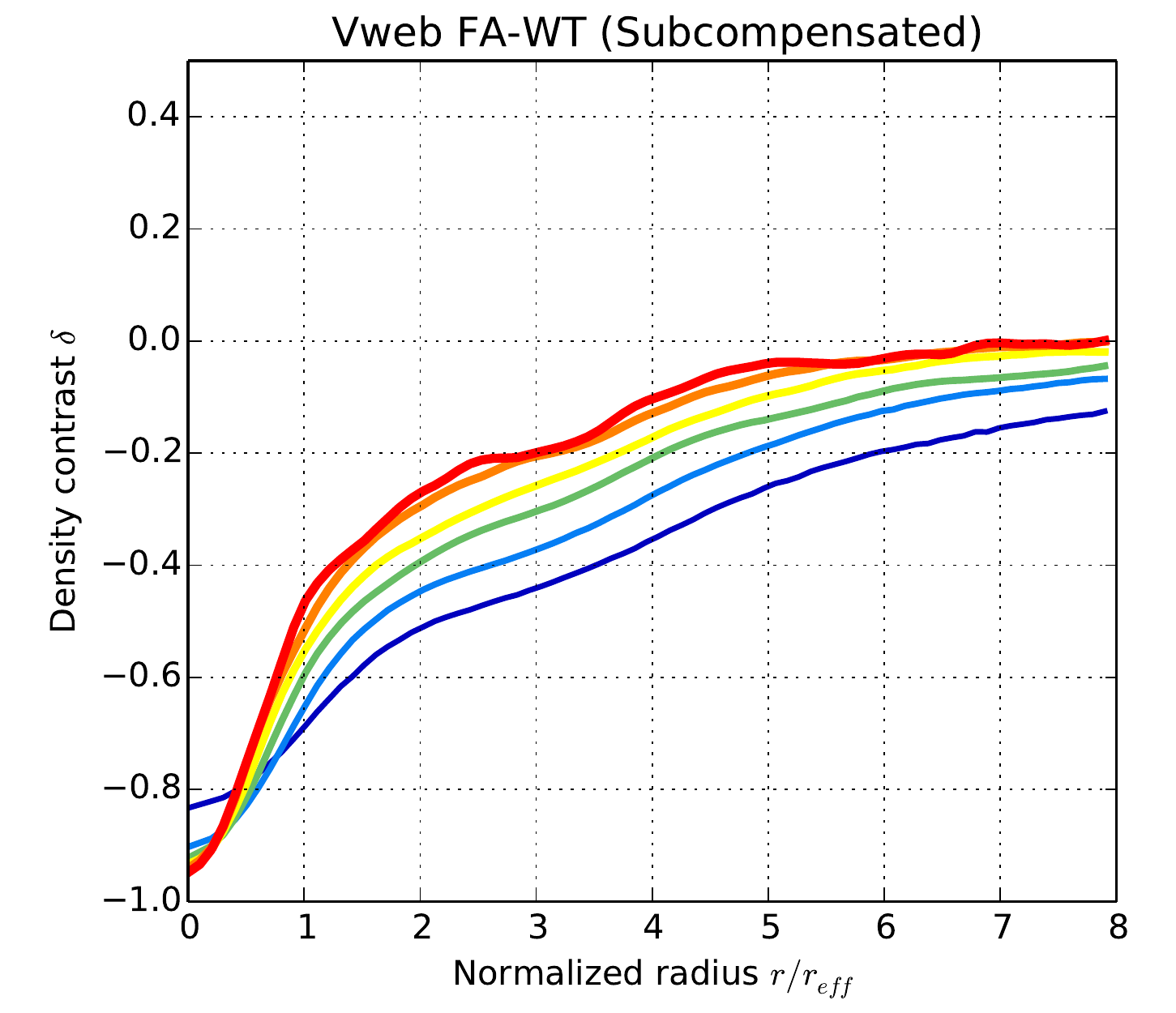}
  \includegraphics[trim = 2mm 2mm 5mm 0mm, clip, keepaspectratio=true,
  width=0.35\textheight]{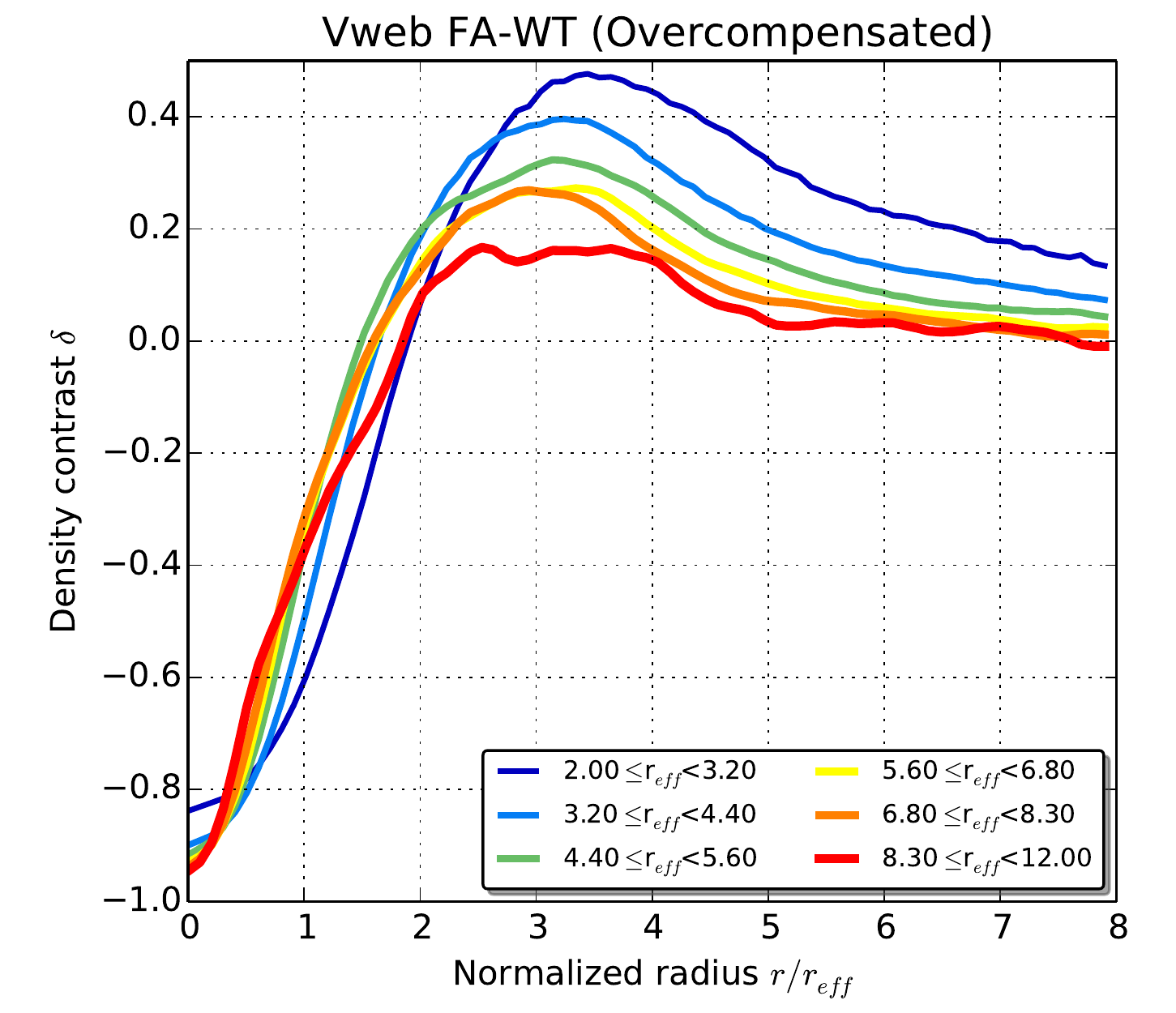}

  \captionof{figure}{\small Spherically averaged radial density profiles for
    stacked voids with different effective radii. The sample is split
    into subcompensated and overcompensated voids (right and left) for
    the web schemes T-Web and V-Web (top and bottom).}
  \label{fig:density_profile}
  \vspace{0.1 cm}
\end{figure*}

\begin{figure*}
\centering  
  \includegraphics[trim = 2mm 2mm 5mm 0mm, clip, keepaspectratio=true,
  width=0.35\textheight]{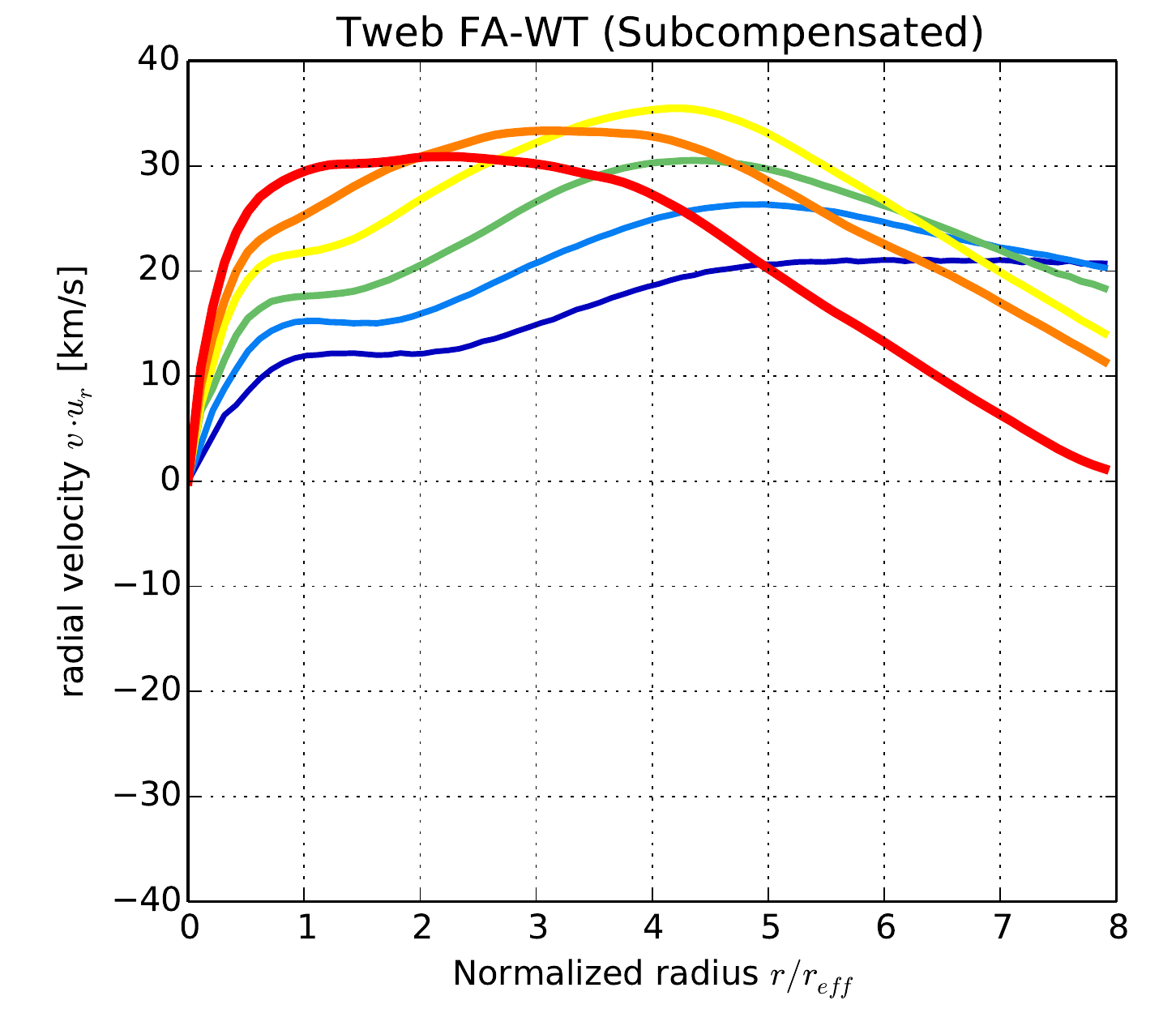}
  \includegraphics[trim = 2mm 2mm 5mm 0mm, clip, keepaspectratio=true,
  width=0.35\textheight]{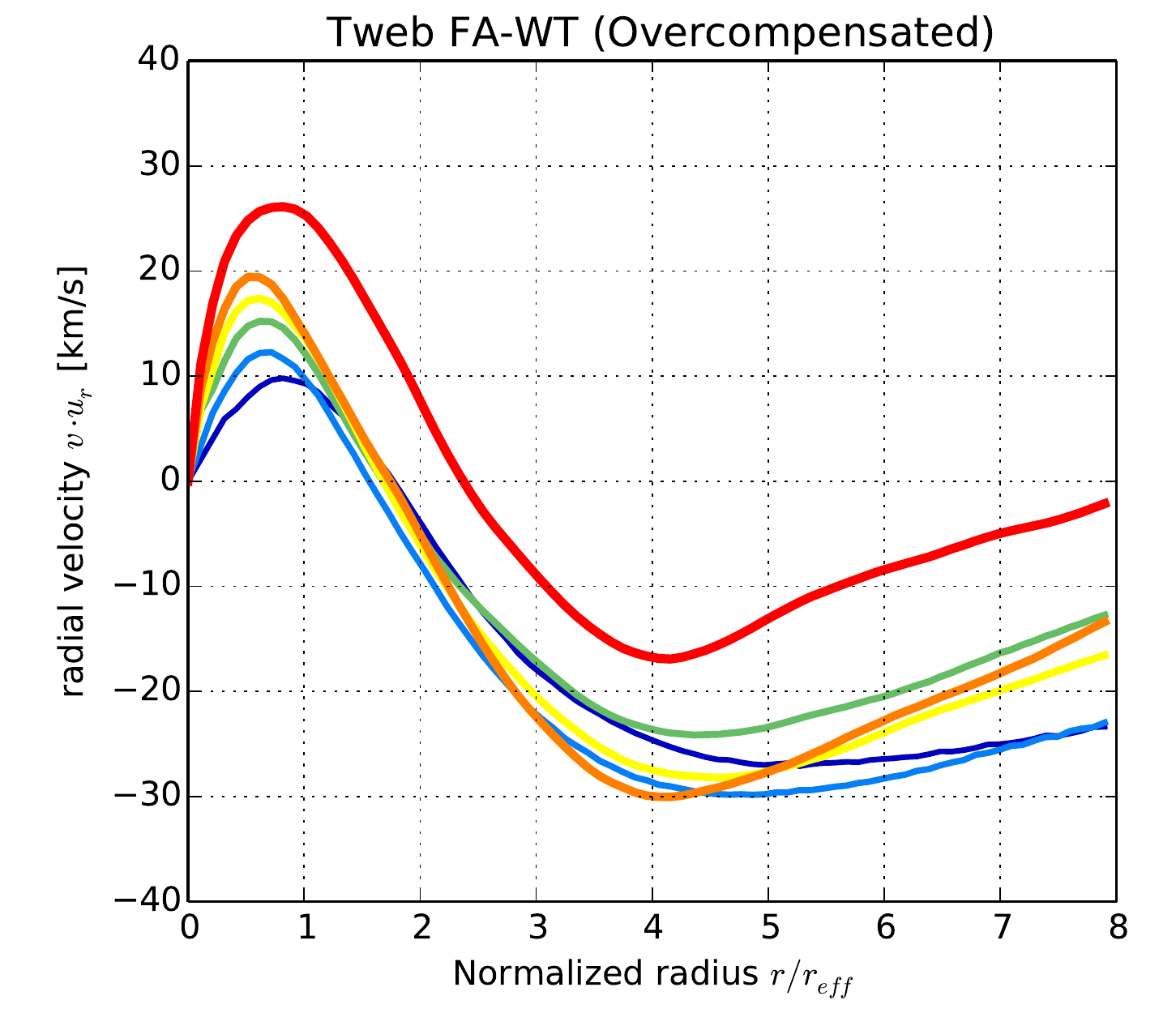}
  \includegraphics[trim = 2mm 2mm 5mm 0mm, clip, keepaspectratio=true,
  width=0.35\textheight]{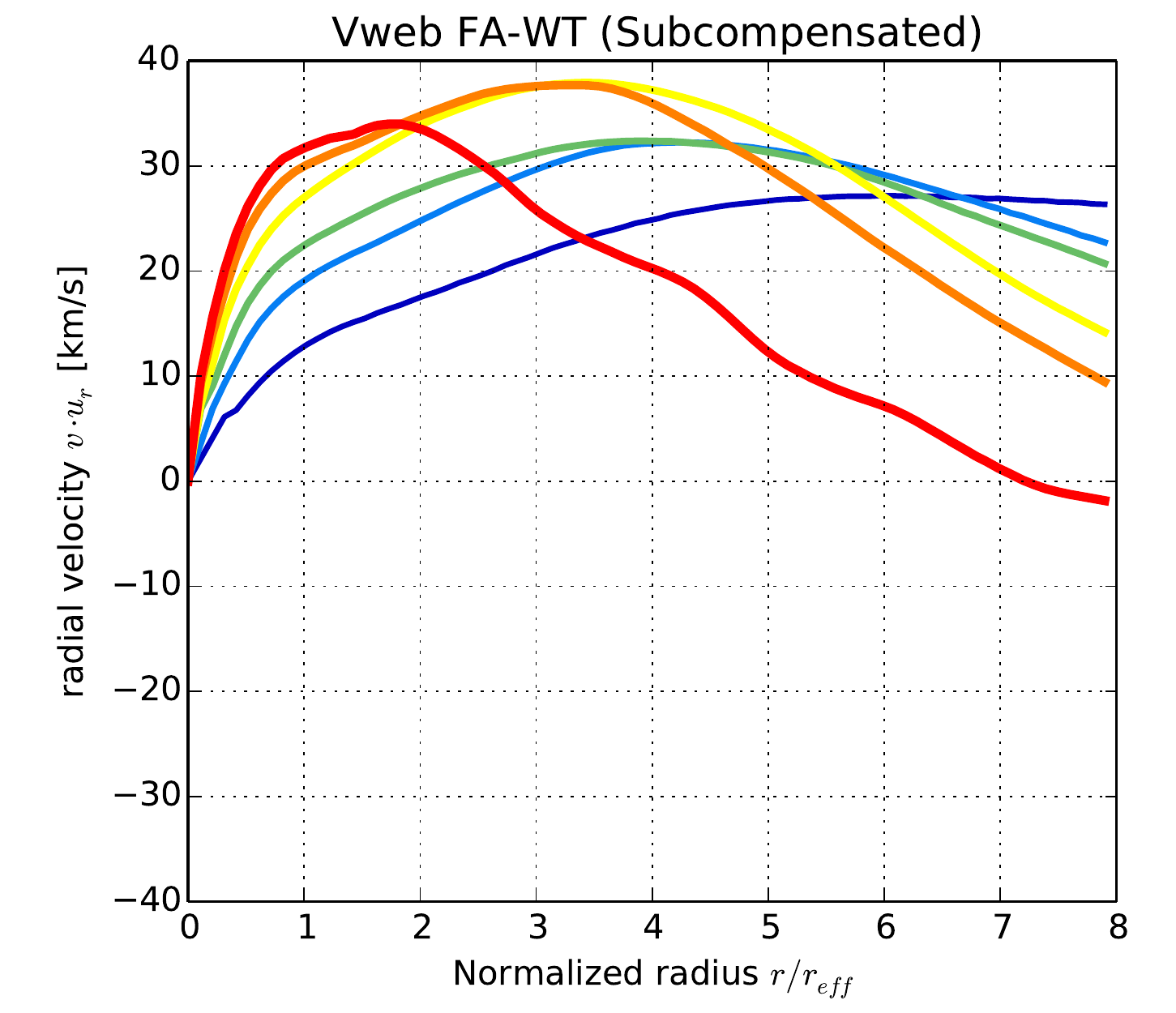}
  \includegraphics[trim = 2mm 2mm 5mm 0mm, clip, keepaspectratio=true,
  width=0.35\textheight]{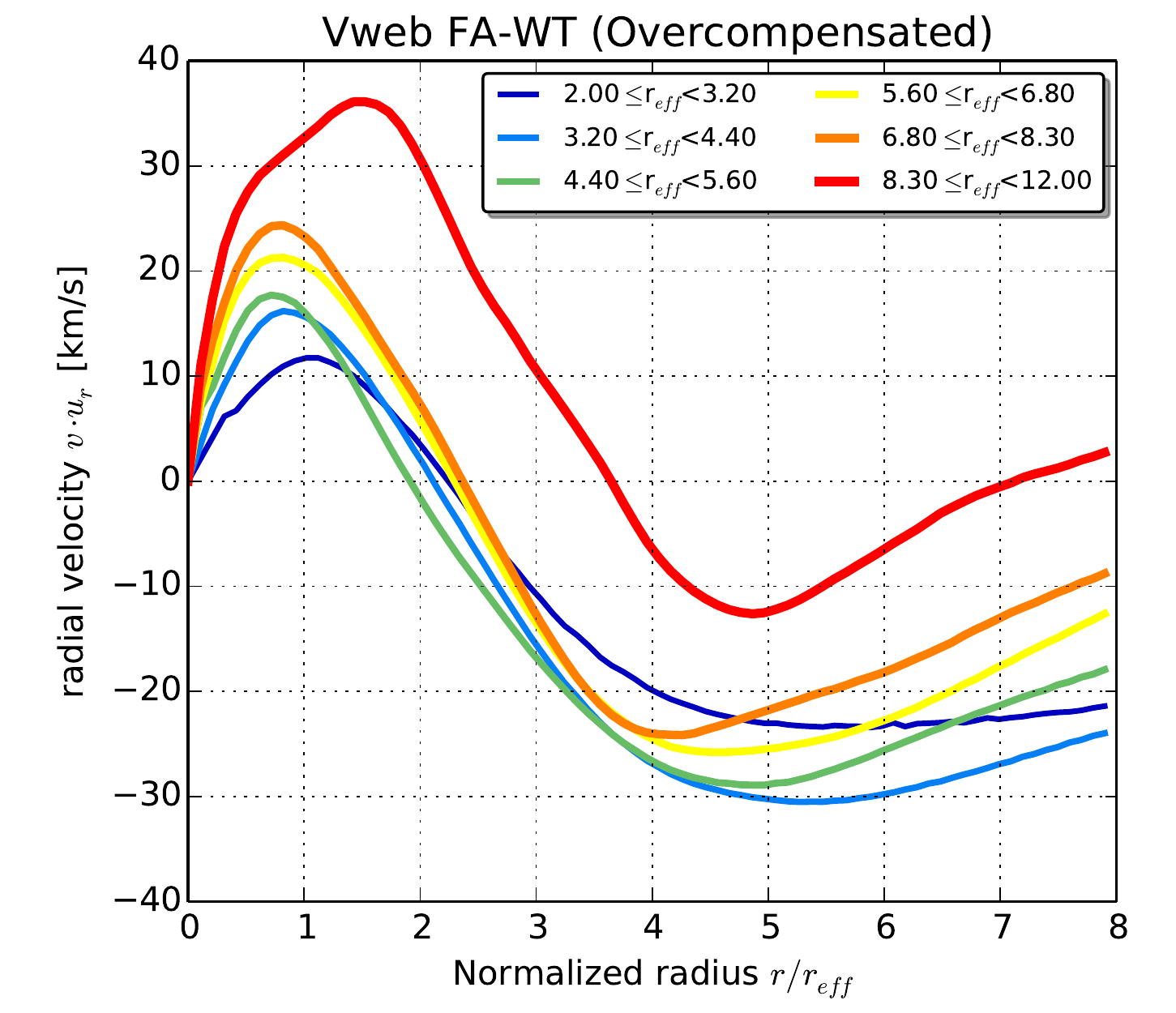}
  \captionof{figure}{\small Spherically averaged radial velocity
    profiles for stacked voids with different effective radii. The
    sample is split into subcompensated and overcompensated voids
    (right and left) for the web schemes T-Web and V-Web (top and
    bottom).} 
  \label{fig:velocity_profile}
  \vspace{0.1 cm}
\end{figure*}

\begin{figure*}
\centering  
  \includegraphics[trim = 2mm 2mm 5mm 0mm, clip, keepaspectratio=true,
  width=0.35\textheight]{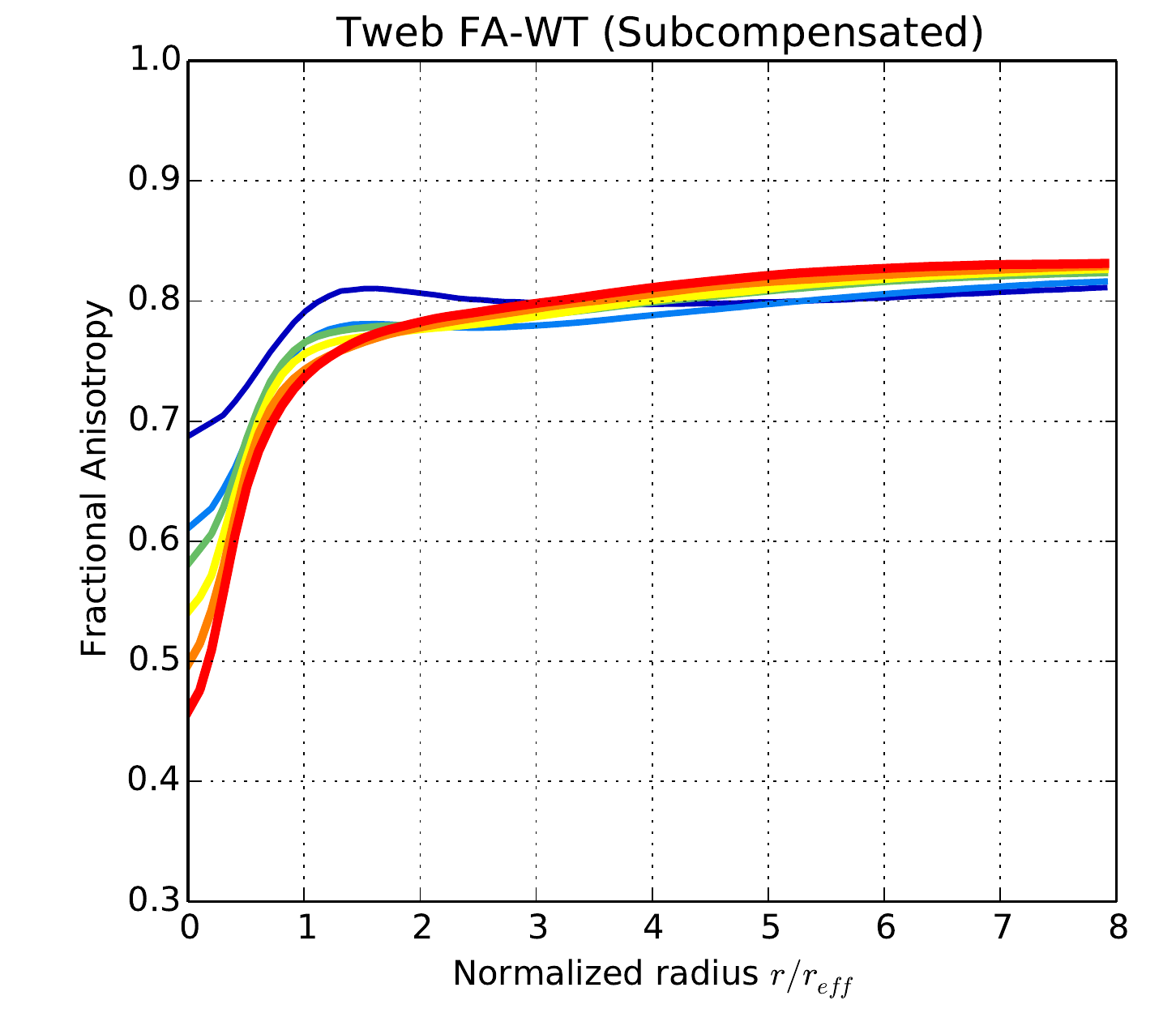}
  \includegraphics[trim = 2mm 2mm 5mm 0mm, clip, keepaspectratio=true,
  width=0.35\textheight]{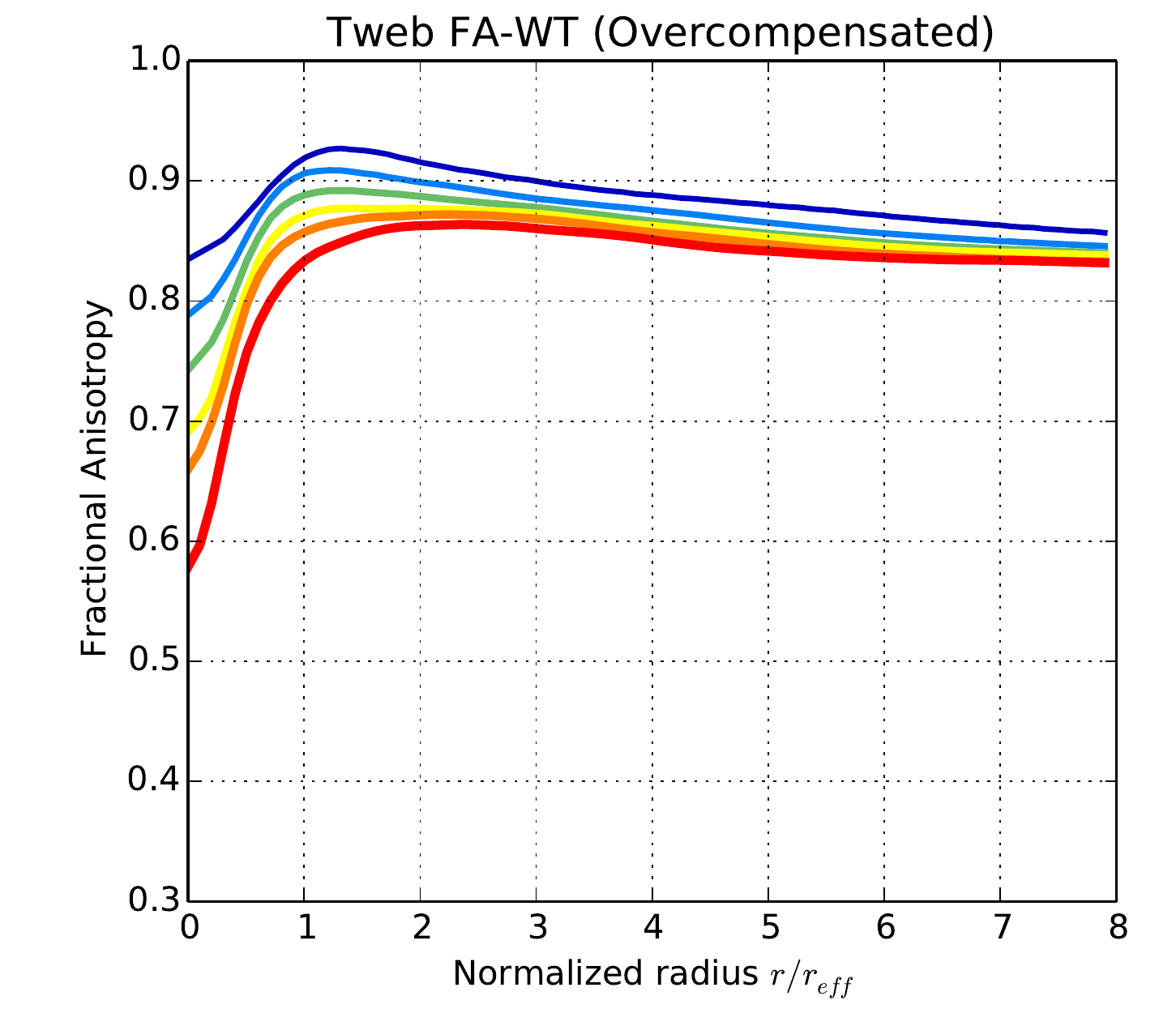}
  \includegraphics[trim = 2mm 2mm 5mm 0mm, clip, keepaspectratio=true,
  width=0.35\textheight]{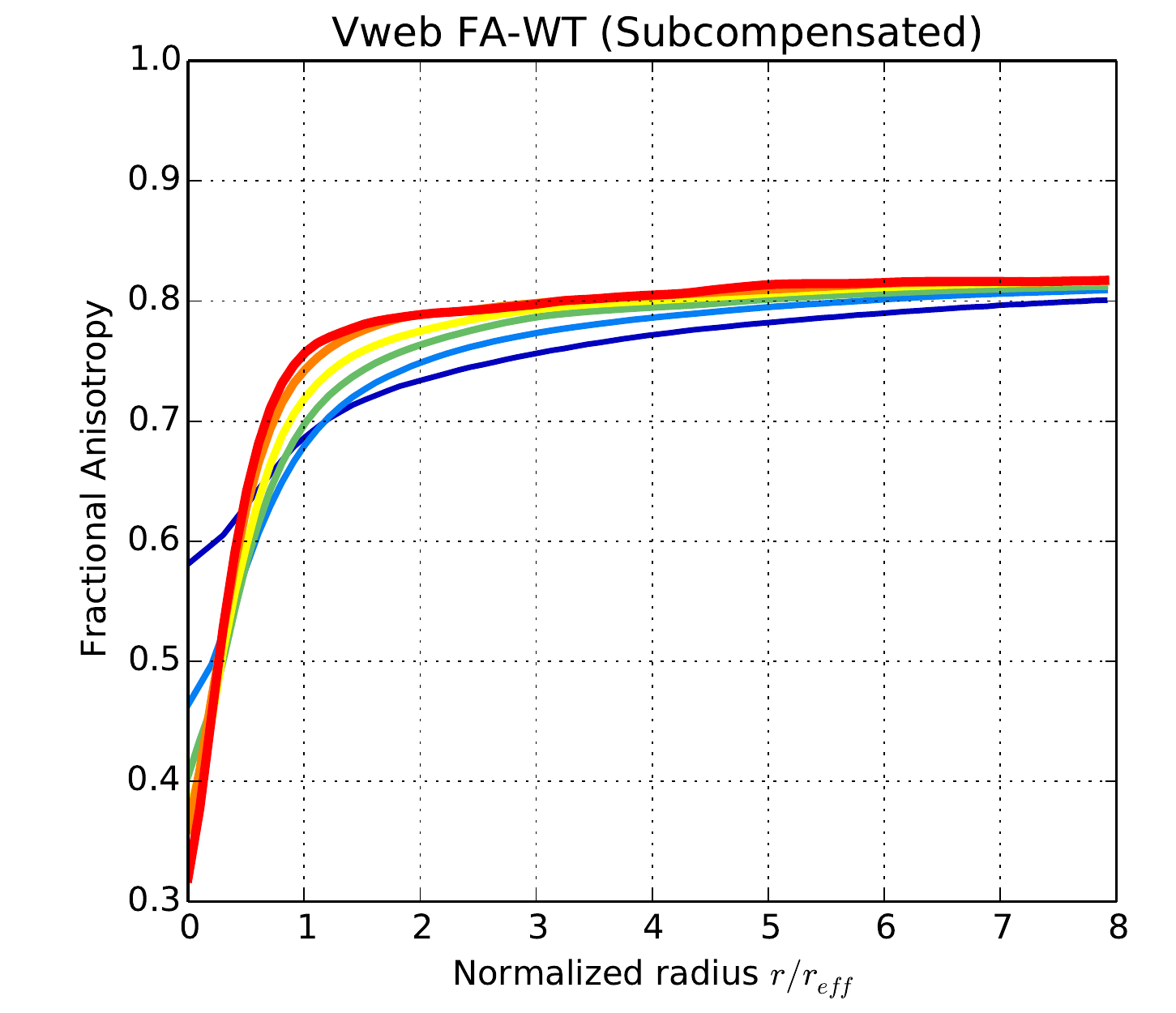}
  \includegraphics[trim = 2mm 2mm 5mm 0mm, clip, keepaspectratio=true,
  width=0.35\textheight]{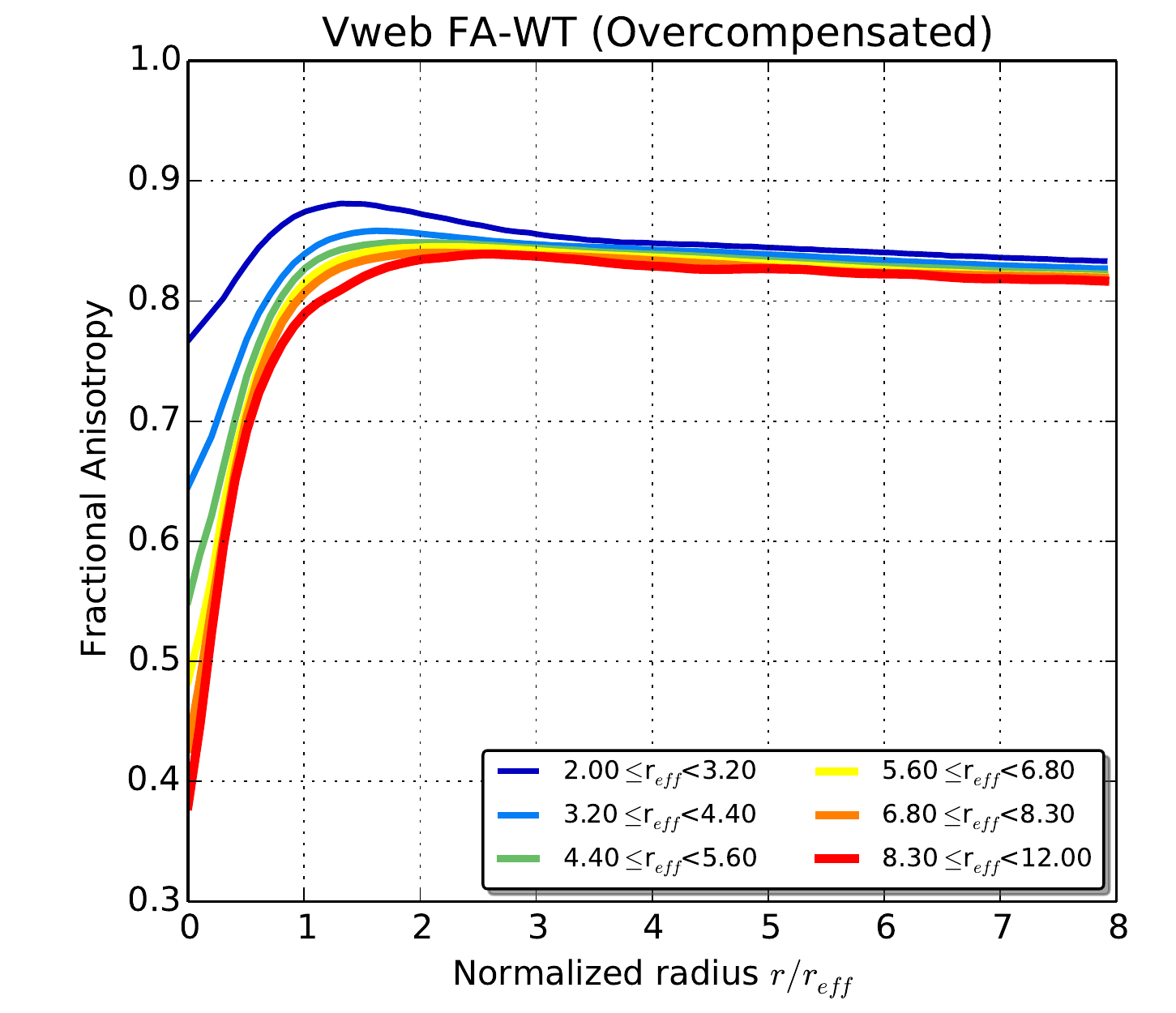}  
  
  \captionof{figure}{\small Spherically averaged Fractional Anisotropy
    profiles for stacked voids with different effective radii. The
    sample is split into subcompensated and overcompensated voids
    (right and left) for the web schemes T-Web and V-Web (top and
    bottom).} 
  \label{fig:FA_profile}
  \vspace{0.1 cm}

\end{figure*}
%.........................................................................

%-------------------------------------------------------------------------
\subsection{Subcompensated and overcompensated voids}
\label{subsec:enclosedmass}
%-------------------------------------------------------------------------

We find that voids are distributed in two different families
differentiated by the 
presence/absence of an overdense matter ridge in their density
profiles. 
To discriminate each void in one of the families we use the
compensation  index $C$
It is defined as the mass of a void enclosed in a spherical volume of
radius $R$ and normalized by the mass of the same volume assuming
it is filled by matter with the mean background density.

\eq{compensation}
{C= \frac{M_v}{\overline{M}} = \frac{3}{2R^{3}} \int_0^{R} [\delta(r) + 1] r^2 dr}

We choose an integration radius of $R=4\reff$, that is large enough to 
enclose the compensation ridge for a typical void in case there is one. 
This leads us to voids with $C>1$ having more mass than expected, 
constituting the family of overcompensated voids. 
These voids generally  exhibit a compensation ridge associated to
dense nearby structures. 
In the same fashion, voids with $C<1$ constitute the family
of  subcompensated voids. 

In Fig. \ref{fig:density_profile} we show the density and velocity profiles
of voids split in these two families. 
In the left column it becomes clear the difference between sub- and
overcompensated voids. 
Table \ref{tab:number_voids} lists the number of voids in 
each effective radius bin for the two web finding methods.

%.........................................................................
\begin{table}[!htbp]
\centering
\begin{tabular}{c | c c | c c}
\toprule
\multicolumn{1}{c}{}&  \multicolumn{2}{c}{\textbf{Tweb}} & \multicolumn{2}{c}{\textbf{Vweb}}\\
%\midrule
\textbf{${\bf \reff }$[\hMpc]}   & $\bf{C<1}$   & $\bf{C>1}$    & $\bf{C<1}$   & $\bf{C>1}$\\ \hline
%\midrule
$<2.0$        &  997 & 495   & 13480  & 711\\
$2.0-3.2$     &  791 & 471   & 2581   & 583\\
$3.2-4.4$     &  941 & 469   & 2168   & 656\\
$4.4-5.6$     &  869 & 467   & 1934   & 724\\
$5.6-6.8$     &  769 & 395   & 1287   & 584\\
$6.8-8.3$     &  665 & 362   & 700    & 315\\
$8.3-12.0$    &  587 & 333   & 123    & 61\\
$>12.0$       &  29  & 23    & 0      & 0\\ \hline
\textbf{total}& 5648 & 3015  & 22273  & 3634\\
\bottomrule
\end{tabular}
\caption{Number of voids for each bin of effective radius and for 
subcompensanted and overcompansated samples.}
\label{tab:number_voids}
\end{table}
%.........................................................................

%.........................................................................
\subsection{Density profiles}
\label{subsec:density_voids}
%.........................................................................

We calculate the contrast density, radial-projected velocity and 
FA profiles. 
For this purpose we stack voids with similar effective radius.
We also compute the profiles separately  for subcompensated and
overcompensated voids. 
For each void,  
we take the distance of each member cell to the void centre along with the 
properties of interest. 
We normalize the radial distance with the effective radius and add the
result to the stack.

Fig. \ref{fig:density_profile} shows the results of stacked density
profiles for different void sizes.  
We calculate the profile out to a radius $8\ \reff$  to capture the
point where the overdensity reaches the mean value.

The first interesting result is the overdensity value at the void's
centre.
We find that larger voids have a lower overdensity value.
The largest voids ($8.3-12\ \hMpc$) have an underdensity
$\delta \sim-0.95$ while smaller voids  ($2-3.2\ \hMpc$) fall around $\delta\sim 
-0.8$ at their centres.
This holds for both web schemes. 
These values are consistent with most of the void finding schemes
based on smooth and continuous fields from simulation or
reconstruction procedures on surveys \citep{Plionis02, Colberg05,
  Shandarin06,  Platen07, Neyrinck08, MunozCuartas11, Ceccarelli13,
  Paz13, Neyrinck13, Ricciardelli2013}, unlike geometrical approaches based 
on point distributions, where central density values are generally higher
\citep{Colberg08}.

A second feature about these profiles is their steepness at inner
regions.  
In subcompensated voids, larger voids are steeper.
Smaller voids exhibit moderate slopes, reaching the mean density at
larger radii than larger voids.
This suggests that small subcompensated voids are embedded into low
density structures, while large subcompensated voids are surrounded
by dense structures, reaching the mean density at lower effective
radii than smaller voids. 
In the overcompensated case larger voids reach first  both the
compensation ridge and then the mean density value.

Regarding overcompensated voids, a final result is related to the height 
of the compensation ridge: the larger the void size, the lower the ridge
height. This implies that overcompensated smaller voids are embedded in 
very high density regions, unlike their subcompensated counterpart, thus 
indicating two possibly different processes for small voids formation. 
Larger voids exhibit lower ridges as outer radial layers also includes all
sort of structures, thus being the difference between large overcompensated
and subcompensated voids less conclusive.

All the previous results hold for both finding schemes.
This suggests an universal behaviour for the radial density profile in
two families of subcompensated and overcompensanted voids. 
This goes in the same direction of recent results about the internal
\citep{Colberg05,  Ricciardelli2013} and external structure of voids
\citep{Lavaux12, Ceccarelli13, Paz13, Hamaus14}. 
Indeed, our results extend the findings of \cite{Hamaus14} into the
range of voids with size $\reff <10$\hMpc.  
 
%.........................................................................
\subsection{Velocity profiles}
\label{subsec:velocity_voids}
%.........................................................................

In Fig. \ref{fig:velocity_profile} we present the radial velocity
profiles. 
Positive values correspond to outflows with respect to the centre.

We find that subcompensated voids have outflowing velocity profiles
all the way up to the radius where the average radial
density reaches zero.
In voids with sizes $\reff <8\hMpc$ the outflow is always
positive, consistent with the fact that their density profiles do not
reach the level $\delta=0$ in the range of explored radii.
This behaviour indicates that matter is being pulled out of the void
into external higher density features.

On the other hand, overcompensated voids initially exhibit outward
profiles, (as expected from a low density region) and approximately at
the radius of the compensation ridge, the velocity reaches a peak,
decreases and becomes negative, showing the infalling flow of matter
further than the compensation ridge. 
This shows that the high density structures associated to the
compensation ridge dominate the matter flow both from inside and
outside the void.

As in the density, these results are also consistent with a Universal
velocity profile for voids \citep{Paz13, Hamaus14}.

%.........................................................................
\subsection{Fractional Anisotropy profiles}
\label{subsec:FA_voids}
%.........................................................................

Fig. \ref{fig:FA_profile} shows the results for the FA profiles.
We find that the FA clearly magnifies the difference between the
internal profile $r/\reff <1$ and the external profile $r/\reff >1$. 
Subcompensanted voids in the T-Web reach an asymptotic background
value of F=$0.8$ almost right at $r/\reff =2$. 
For the V-Web the results are similar, albeit with a slower trend with
the effective radius. 
Overcompensated voids reach a maximum FA at the same effective radius
$r/\reff =1$ and decline to reach an asymptotic value of FA=$0.85$
at larger radii.

The central FA values also show a magnified trend with the void size. 
Large voids have lower FA values, amplifying the same trend
observed with the central density value. 
Finally, we observe that voids in the V-Web scheme span a larger range
of central FA values than in the T-Web scheme.  

The difference between the radius where the density ridge is reached 
($r/\reff =3$) and the radius of the FA maxima ($r/\reff =1$, most
manifest in the T-Web voids) justifies in a more quantitative way the
qualitative argument we present in sub-section \ref{subsec:watershed}
to define void boundaries in our method.  
Namely that as we increase the void boundary moving away from the
centre, we reach first middle density walls, before reaching high
density structures associated with high anisotropy.  
This implies that voids defined with the FA never present positive
overdensities, making the FA maximum a reasonable boundary for voids
compared to the traditional definition that puts the boundary at the
density ridge.

We recognize that our choice produces smaller voids as compared with
other voids finding methods. 
We consider that this has the advantage of
avoiding the contamination from external structures.

%*************************************************************************
\section{Conclusions}
\label{sec:conclusions}
%*************************************************************************

In this paper we propose that the anisotropy of the eigenvalues from
tidal and velocity shear tensors is a good tracer of  cosmic voids. 
Based on this idea we go on to implement a watershed
algorithm on the fractional anisotropy to find voids in a N-body
cosmological simulation.
We perform different test on the voids found on the anisotropy of two
tensorial schemes, the T-Web and the V-web.  

The first quantity we examine is the void size distribution
characterized by an effective radius. 
We find that for the T-Web scheme the void size distribution has a
shape consistent with standard expectations from a two-barrier setup. 
However, for the V-Web we find an overabundance of the smallest voids.
This is due to two reasons. 
First, there are artefacts in the web
finding scheme inside sheets. 
Second, there V-web produces smaller
fragmented voids inside the largest voids found in the T-Web
classification.  
This void fragmentation hints at a complex velocity structure in large
voids compared to its simpler density/gravitational potential anatomy.  

A second step in the void characterization 
is the separation into subcompensated and overcompensated samples. 
In the T-Web $35\%$ of the voids are overcompensated and $65\%$
subcompensated, meaning that they are located in denser regions with
a clear delimiting ridge.   
For the V-web the subcompensated fraction rises to $85\%$ as the
void size decreases, supporting the picture of smaller spurious voids
located in medium density matter sheets.   

Finally we proceed with a more detailed void characterization
through the radially averaged profiles of density, radial velocity and
fractional anisotropy. 
In this case we study separately the subcompensated and
overcompensated voids, split this time in samples with different
effective radii. 

For the density the most interesting feature is that the  profile has
a similar shape for all the different void sizes once the radial
coordinate is expressed in units of the effective radius.  
In these profiles it is evident the presence of an overdense ridge
around $2<r/\reff <3$ for the overcompensated voids.
Subcompensated voids do not show such a ridge keeping its density
lower than the average value up to $r/\reff \sim 8$.

In other studies where voids are found using considerations on the
density fields the ridge is, almost by definition, located around
$r/\reff \sim 1$ and the subcompensated voids reach average
density around $r/\reff \sim 3$. 
This indicates that voids defined by FA boundaries are smaller  than
voids found by density only considerations close by a radial factor of
$2$.   

The velocity profiles show the expected correlation with the features
already observed in the density.
Most notably the radial velocity goes to zero close to flat regions in
the density profile and is positive around regions of increasing
density with radius.  

The FA profiles also show remarkable similarities
inside each class of subcompensated and overcompensated voids
and a close correlation with the behaviour in the density and velocity
profiles. 
From these profiles it becomes clear that the limit of overcompensated
void coincides with a ridge in of FA values close to $0.9$. 
This value is close to the limit FA=$0.95$ we impose in the watershed
algorithm, indicating that these voids are very close to spherical.
For the subcompensated voids the FA quickly increases to FA values of 
$\sim 0.7-0.8$ at $r/\reff \sim 1$ to find a plateau, which is
quite far from the limiting FA$=0.95$ suggesting that these voids are
ellipsoidal. 

Put together, the results for the radially averaged profiles support the 
evidence for a universal density profile.  
Our results extend to smaller void sizes the work done by
\cite{Hamaus14} on simulations and by \cite{Ceccarelli13} on SDSS data.

The method we present here finds voids with reasonable properties
compared to different results published in the literature. 
This gives support to a new tracer, the Fractional Anisotropy, to be
used in the study of cosmic voids. 
Furthermore, the complementary physical picture in the two different
web finding algorithms opens the possibilities to make a joint
analysis of the tidal and dynamical structure of voids.

%*************************************************************************
\section*{Acknowledgments}  
%*************************************************************************

We would like to thank Fiona Hoyle and Rien van de Weygaert for their
exhaustive and thoughtful reviews, accompanied with very detailed and
helpful suggestions on earlier versions of this manuscript.  
We also thank Nelson Padilla and Yehuda Hoffman for enlightening
comments and discussions.
SB also thanks Juan Carlos Mu\~noz-Cuartas 
for many clarifying discussions and helpful ideas.
JEFR acknowledges financial support from Vicerrector\'ia de
Investigaciones at Universidad de los Andes (Colombia) through a FAPA
grant. 

\bibliographystyle{mn2e}
%\bibliography{references}

\end{document}